\documentclass[twocolumn,amsmath,amssymb,floatfix,pra,12pts,graphicx]{revtex4-2}
\usepackage[caption=false]{subfig}  

\usepackage{overpic} 
\usepackage[utf8]{inputenc}
\usepackage{epstopdf}
\usepackage{subfig}

\usepackage{dcolumn}
\usepackage{bm}
\usepackage[mathlines]{lineno}
\usepackage{xcolor}
\usepackage{soul}
\usepackage{bm}
\usepackage{bbm}
\usepackage{amsmath,amsfonts,amssymb}
\usepackage{extarrows} 
\usepackage{mathtools}
\usepackage{enumitem} 
\usepackage{algorithmicx}
\usepackage{algpseudocode}
\usepackage{mdframed}
\usepackage{xr}
\usepackage{xcolor}

\usepackage{hyperref}
\hypersetup{
colorlinks,citecolor=red,
filecolor=blue,
linkcolor=blue,
urlcolor=black
}

\definecolor{sph}{rgb}{0.0588, 0.3216, 0.7294} 
\definecolor{ppk}{rgb}{1.0, 0.4549, 0.0902} 
\newcommand{\nc}{\newcommand}
\nc{\nn}{\nonumber}
\nc{\txt}{\textrm}
\newcommand{\beq}{ \begin{equation} }
\newcommand{\eeq}{ \end{equation} }
\nc{\txtsup}{\textsuperscript}
\nc{\txtsub}{\textsubscript}
\nc{\calL}{\mathcal{L}}
\nc{\U}{\mathcal{U}}
\nc{\T}{\mathcal{T}}
\nc{\E}{\mathcal{E}}
\nc{\cH}{\mathcal{H}}
\nc{\sect}[1]{{\sl #1 .--}}

\newcommand\nya[1]{\{ #1 \}}

\nc{\RV}[1]{\textcolor{blue}{#1}}
\nc{\YD}[1]{\textcolor{red}{#1}}

\nc{\cC}{\mathcal{C}}
\nc{\cI}{\mathcal{I}}

\begin{document}
\title{
Qlustering for Data Clustering via Network-Based Quantum Transport 
}

\author{Shmuel Lorber and Yonatan Dubi}
\affiliation{Department of Chemistry, Ben Gurion University of the Negev, 1 Ben-Gurion Ave, Beer Sheva 8410501, Israel }
\email{shmuell@post.bgu.ac.il; jdubi@bgu.ac.il}
\date{\today}
\begin{abstract}
Analog quantum computation offers a route to machine learning using controllable physical dynamics as a computational resource. However, many existing approaches rely on task-specific protocols or observables that are difficult to access experimentally, limiting generality and implementation. Here we introduce Qlustering, an unsupervised clustering framework based on steady-state quantum transport in quantum networks governed by the GKSL master equation, developed through algorithm-hardware co-design. Data are encoded as input states, and cluster assignments are inferred from steady-state output currents, avoiding full state tomography in favor of accessible transport observables. The method realizes a hybrid classical-quantum workflow in which data preparation and training are performed classically, while clustering is carried out by transport dynamics. We benchmark the method on synthetic datasets, localization, and QM9 and Iris, finding competitive performance and stability over a broad range of dephasing strengths. These results show that unlabeled data structure can be extracted directly from steady-state transport observables, identifying terminal-current readout as a native, tomography-free mechanism for unsupervised learning in open quantum networks.
\end{abstract}
   
\maketitle

\section{INTRODUCTION}

Quantum transport in engineered systems provides a physically grounded setting for information processing based on continuous-time dynamics. In analog quantum platforms, computation can be encoded directly in Hamiltonian evolution and open-system dynamics, allowing interference and dissipation to act as computational resources \cite{PRXQuantum.2.017003,Daley2022}. Such dynamics naturally support tasks that can be formulated in terms of structured exploration of state space, including optimization and feature extraction \cite{Biamonte2017,Das2008}.

A central limitation of current transport-based and analog quantum approaches is their restricted scope. Many such schemes are tailored to specific tasks, encodings, or readout protocols, which can limit their generality as learning architectures. A further challenge is their reliance on observables that are difficult to access experimentally, such as the full quantum state or its detailed time dependence. State tomography and transient measurements impose substantial overhead and generally scale poorly with system size \cite{Guo2024CommsPhys,Wang2024SciAdv}. Moreover, coherent transport signatures are often short-lived and sensitive to decoherence, so transient dynamics do not by themselves provide a stable or trainable basis for structure extraction \cite{Albash2018,preskill2018quantum}.

Recent work has shown that steady-state quantum transport networks provide a natural framework for machine-learning tasks in which output currents serve as the primary computational observable. In supervised settings, such networks have been used for quantum state discrimination and classification, demonstrating that steady-state currents can encode task-relevant information without requiring full state reconstruction \cite{dalla2020quantum,wang2022implementation,lorber2024using}. The basic intuition is that an input state injected into a structured transport network undergoes scattering, interference, and dissipation before reaching the output terminals. The resulting current vector therefore acts as a transport-induced embedding of the input: data points with similar underlying structure are expected to generate correlated current patterns, while dissimilar inputs are routed toward different output signatures. In this sense, the network functions as a physical filter that converts latent structure in the data into measurable terminal currents. Figure~\ref{figure 2} illustrates this architecture, where an input state is injected at the source, propagates through the network, and is assigned according to the output port carrying the largest steady-state current. Building on our previous work on supervised current-based classification \cite{lorber2024using}, the present work asks whether the same steady-state transport mechanism can also reveal structure in unlabeled data, thereby enabling clustering through current readout alone.
At the same time, supervised classification alone does not establish that the underlying transport dynamics define a broader learning framework, since classification can be implemented in many classes of tunable networks. Unsupervised classification of quantum data poses a distinct question: rather than learning a label map supplied externally, it asks whether structure can emerge directly from the data themselves. For quantum states, this distinction is especially meaningful because similarity is not fixed a priori, but depends on physical notions such as overlap, distinguishability, or measurement statistics \cite{sentis2019unsupervised}. Unsupervised learning therefore probes whether the dynamics induce a meaningful organization of the data, rather than merely fitting a prescribed mapping \cite{gujju2024quantum,lazarev2025hybrid,yu2022experimental}.

In the present work, we introduce \emph{Qlustering}, a clustering scheme based on steady-state currents generated by open-system quantum dynamics governed by the GKSL master equation. The central contribution is not a new Hamiltonian-level transport mechanism, nor a claim of generic superiority over classical clustering algorithms, but the identification of steady-state transport currents as a native, experimentally accessible representation for unsupervised structure extraction. In contrast to supervised current-based classification, where the network learns a prescribed label map, Qlustering tests whether unlabeled inputs can self-organize through the transport dynamics themselves. In this sense, the output-current vector is not merely a readout channel, but a physically generated embedding whose correlations define the cluster structure.

Data are encoded as input states of a tight-binding Hamiltonian, and cluster assignments are inferred from correlations in the steady-state currents measured at designated output terminals. By optimizing the Hamiltonian parameters with respect to an unsupervised objective, the dissipative transport dynamics define a nonlinear embedding in current space. Inputs with related structure can then generate correlated transport signatures, enabling cluster formation without explicit labels and without relying solely on pairwise distance computations.

Performance is evaluated on synthetic benchmarks, a localization task, and representative chemical and biological datasets (QM9 and Iris). Since the present work is intended as a theoretical proof of concept, we do not attempt an exhaustive benchmark against the full landscape of clustering algorithms. Instead, we use k-means as a standard and widely recognized reference method. Across these settings, Qlustering achieves competitive performance relative to this baseline, while relying only on steady-state current measurements. We also examine robustness to dephasing noise and find stable performance over a broad range of noise strengths (see Appendix~\ref{appendix: dephasing}).

As for implementation, photonic platforms provide a natural setting for realizing the required transport dynamics, as discussed in Sec.~\ref{Physical_Implementation}. Caruso \textit{et al.}~\cite{caruso2016fast} demonstrated random quantum walks in a photonic maze architecture closely related to the dynamical equations considered here. Programmable photonic processors~\cite{harris2018linear,harris2017quantum} offer a flexible route to implementing the tight-binding networks underlying Qlustering. More generally, the framework is compatible with analog quantum platforms that support programmable couplings, open-system injection and extraction, and terminal current readout. Although less naturally suited to this setting, present-day superconducting quantum processors may also provide a possible route for implementing Qlustering.

These results identify steady-state quantum transport as a physically grounded open-system learning framework in which directly measurable currents can support both supervised and unsupervised tasks.

\section{SETUP AND FORMULATION}
\label{setup}
\begin{figure} 
\centering
\includegraphics[width=\linewidth, trim={0 130 0 6cm}, clip]{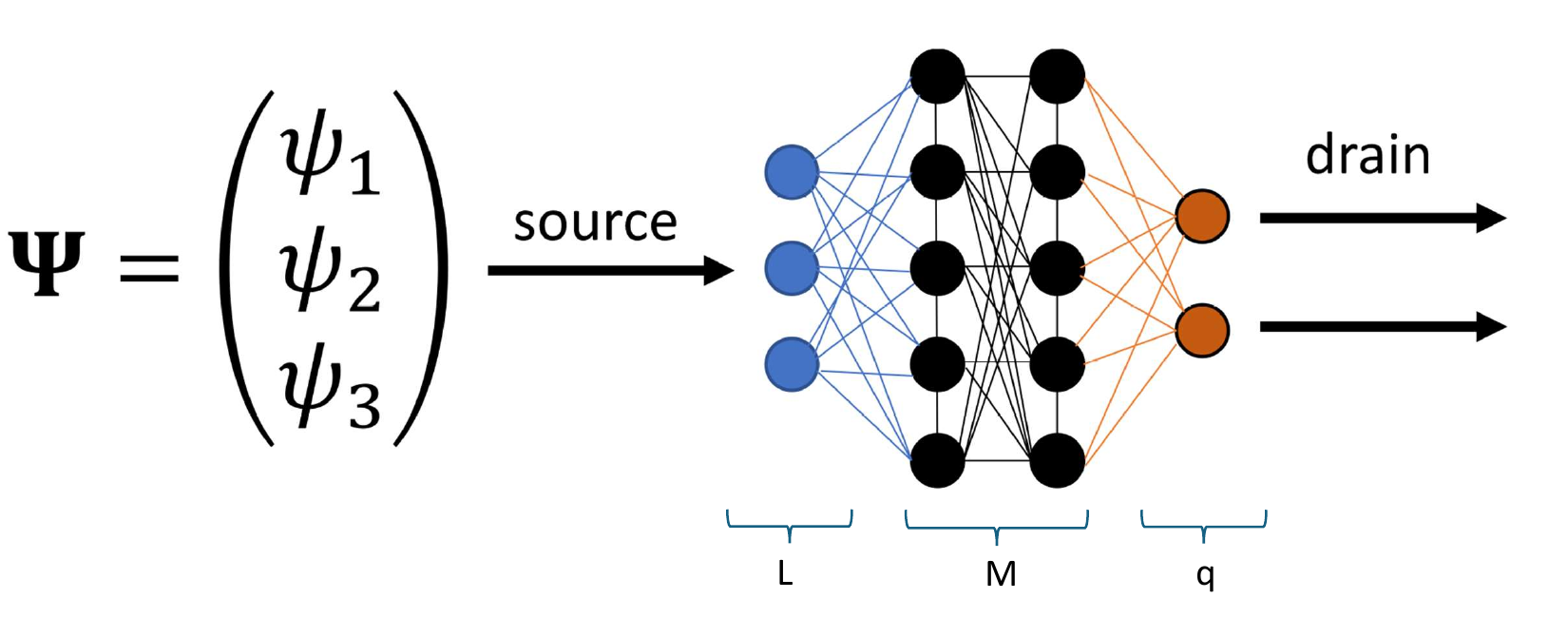}
\vspace{-0.5em} 
\caption{\textbf{Schematic representation of the Qlustering network.}
An input state vector $\Psi$ of dimension $L$ is injected into the network through the input nodes. In this figure, $L=3$. The propagation, injection, and extraction of a particle are modeled by Eq.~\ref{Lindblad}. Here, $L$ denotes the number of input nodes and also the dimensionality of the state vectors, $q$ is the number of output nodes (i.e., clusters), and $M$ is the number of hidden nodes. After reaching a steady state, the current from each output node is computed. The state is then assigned to the group corresponding to the output node with the highest current.}
\label{figure 2}
\end{figure}

\normalsize \textbf{The Qlustering algorithm.--} 
We begin by describing the physical and structural properties of the Qlustering device. Given a set of $N$ state vectors, $\left\{ \Psi_n \right\}$, in an $L$-dimensional Hilbert space, and a predetermined number of clusters $q$, the task is to assign each $\Psi_n$ to its corresponding class $\mathcal{C}[\Psi_n]$. The network is constructed as follows (see Fig.~\ref{figure 2}): consider a system-environment combination in which the system is the network, and the environment acts as a reservoir. The $L$ input nodes and $q$ output nodes are connected to the environment (such that particles can flow into the input nodes and out of the output nodes) but are not directly linked to each other. Between the input and output layers, there are $M$ "hidden" nodes. The dynamics of particles in the network are governed by a tight-binding Hamiltonian of the form $\cH=\sum_{i,j} h_{ij} c^\dagger_i c_j + \text{h.c.}$, where $c^\dagger_i$ ($c_i$) creates (annihilates) a particle at node $i$. The Qlustering algorithm is parameterized by the hopping terms \( h_{ij} \) of \( \mathcal{H} \).

To Qluster a given data element $\Psi_n$, one excites the input nodes (in a manner that encodes $\Psi_n$) and allows the particle to propagate freely within the network until it reaches a steady state. Once steady state is achieved, the output current vector is computed as $J[\Psi_n]=\begin{bmatrix} j_1 \\ j_2 \\ \dots \\ j_q \end{bmatrix}$. The output node $i \in \{1, \dots, q\}$ that yields the highest current $j_{\text{max}}$ determines the class to which $\Psi_n$ is assigned. After processing the entire dataset, clusters are defined by grouping together the state vectors associated with the same output node corresponding to $j_{\text{max}}$. 
The clustering behavior, dictated by the hopping terms $h_{ij}$ of the Hamiltonian $\cH$, is optimized through an iterative procedure as follows: 

\begin{mdframed}[linewidth=0.5pt]
\noindent\textbf{ Qlustering algorithm}\begin{algorithmic}[1]
    \State Initialize the Hamiltonian $\mathcal{H}$ randomly.
    \State Propagate each state vector $\Psi_n$ through the network and calculate its steady-state current $J[\Psi_n]$, and the overall currents $J=J[\Psi_1],J[\Psi_2], \dots, J[\Psi_N]$.
    \State Evaluate the clustering cost function
    \begin{equation}
        \label{CF}
        CF(J, N) = \sum_{n=1}^N (I_n - J_n)^2 
    \end{equation}
    While $J_n=J[\Psi_n]$, and $I_n$ is a one-hot vector of length $q$ defined by 
    $I_n(i) = \delta_{i,i_{\max}(n)}$, with $\delta_{i,j}$ the Kronecker delta, i.e. it is a vector 
   with all entries zero except a single one at the index corresponding to the maximal output current for $\Psi_n$ where $n$ runs over the full dataset
    \State Randomly select an entry in $\cH$ and modify it to produce a new Hamiltonian $\cH^{(2)}$.
    \State Recompute cost; keep $\mathcal{H}^{(2)}$ if it decreases.
    \State Repeat Steps 2–5 until convergence.
\end{algorithmic}
\end{mdframed}
Note that when the Hamiltonian changes, the currents $J_n$ are modified, and consequently the vectors $I_n(i) = \delta_{i,i_{max}(n)}$ and the cost function also change, i.e. we use a dynamical cost function. This contrasts with the classification cost function presented in~\cite{lorber2024using}, which remained fixed throughout the training process.\\
The intuition behind this cost function is that, at every iteration, the input state vectors are allowed to change their class. In simple terms, the cost function in Eq.~\ref{CF} selects Hamiltonians that enforce extreme steady-state currents from a single output, thereby reassigning each input’s class according to its output current. We observe that this decision rule naturally drives similar inputs to propagate toward similar outputs (through lowering of the dynamical cost), steering the system toward future clustering-behavior that does not emerge without enforcing this rule.\\ 
To accelerate convergence, we employ multiple particles in Step~4—that is, we draw several values for the selected entry in $\cH$ and compute, in parallel, the corresponding currents and cost function ($CF$) for each.

\begin{widetext}

\begin{figure}[t]
    \centering
    \captionsetup[subfloat]{labelformat=empty} 

    \begin{minipage}[t]{0.3\linewidth}
        \centering
        \textbf{(a)} \\[0.3em]
        \includegraphics[width=\linewidth]{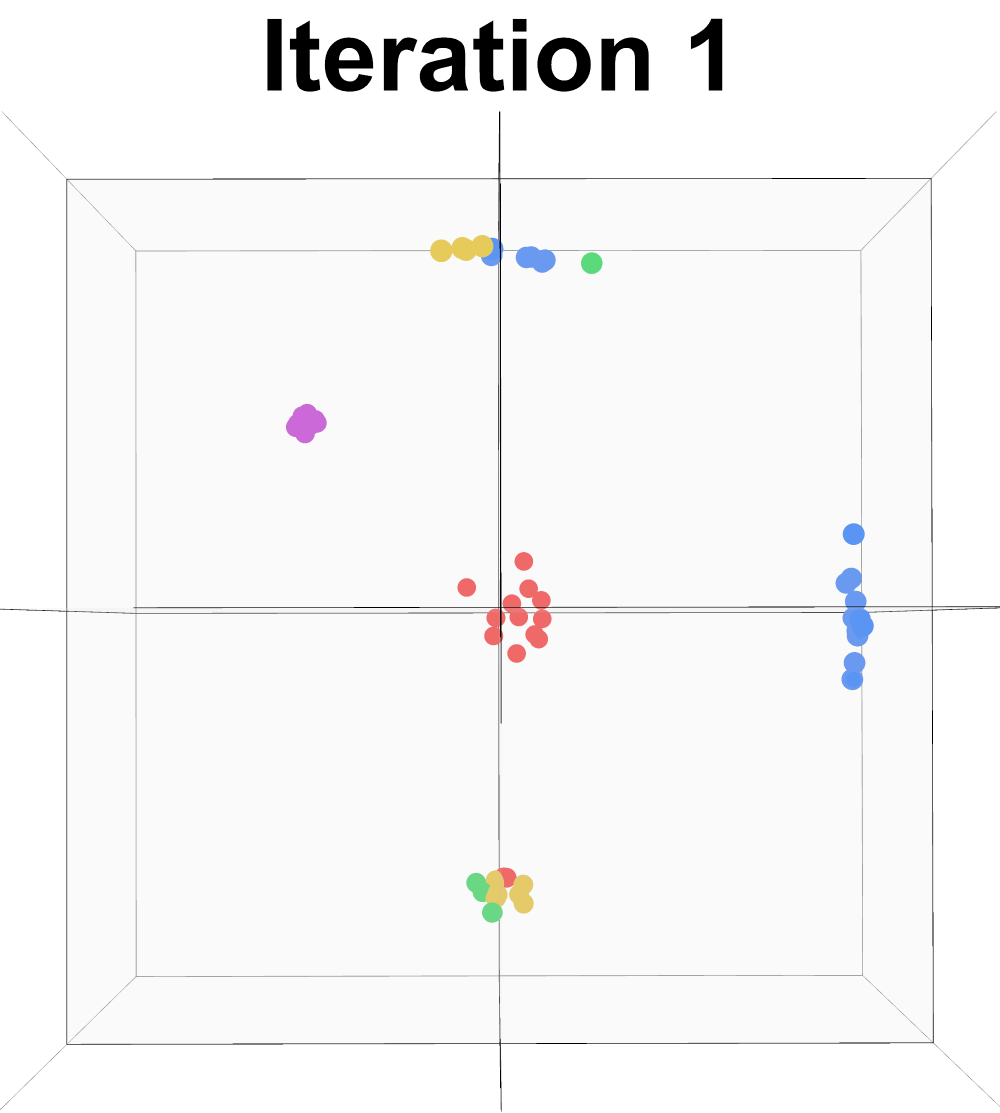}\\[0.5em]
        \includegraphics[width=\linewidth, trim={0 250 0 0cm}, clip]{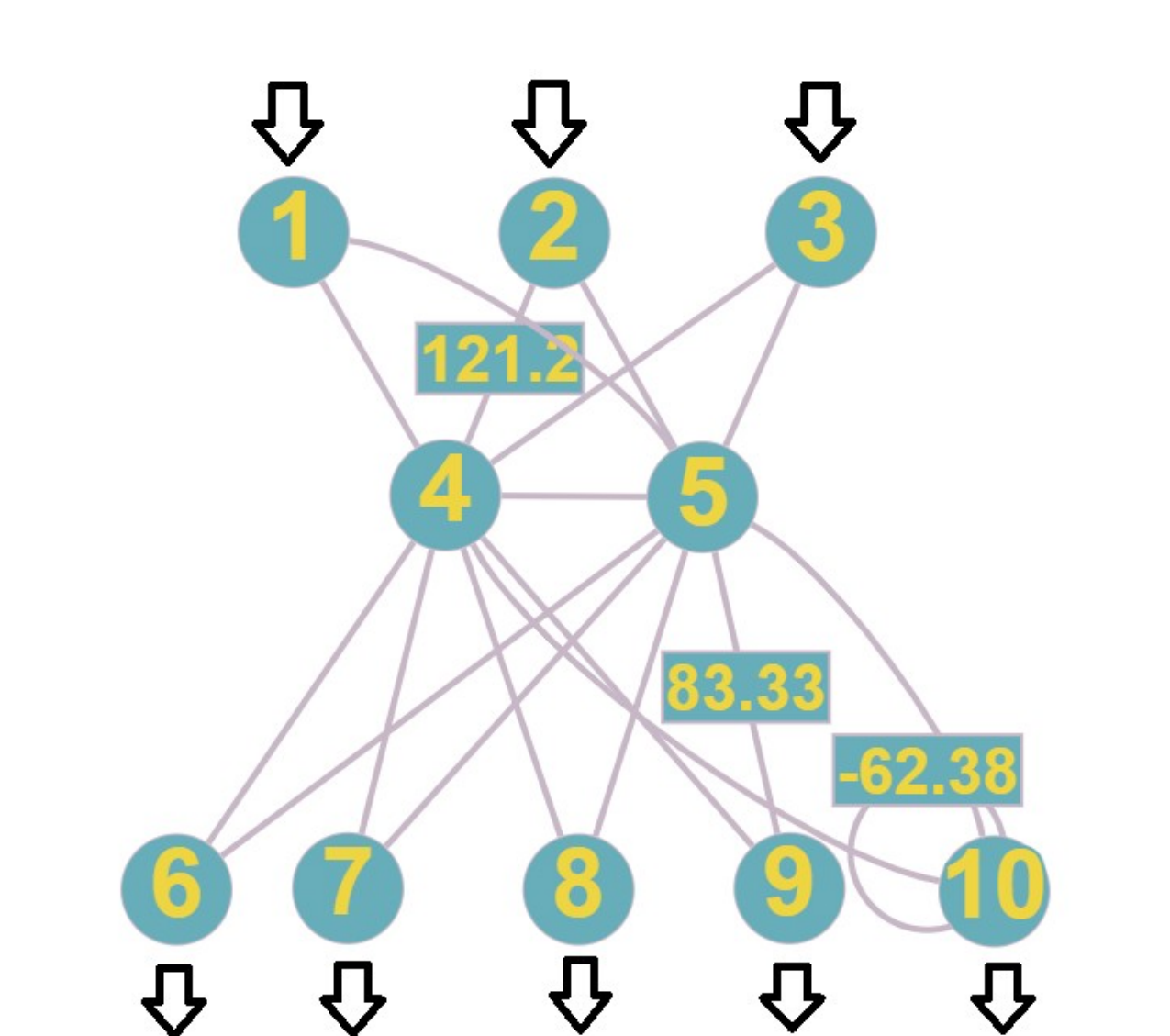}
        \label{fig:col-a}
    \end{minipage}
    \hfill
    \begin{minipage}[t]{0.3\linewidth}
        \centering
        \textbf{(b)} \\[0.3em]
        \includegraphics[width=\linewidth]{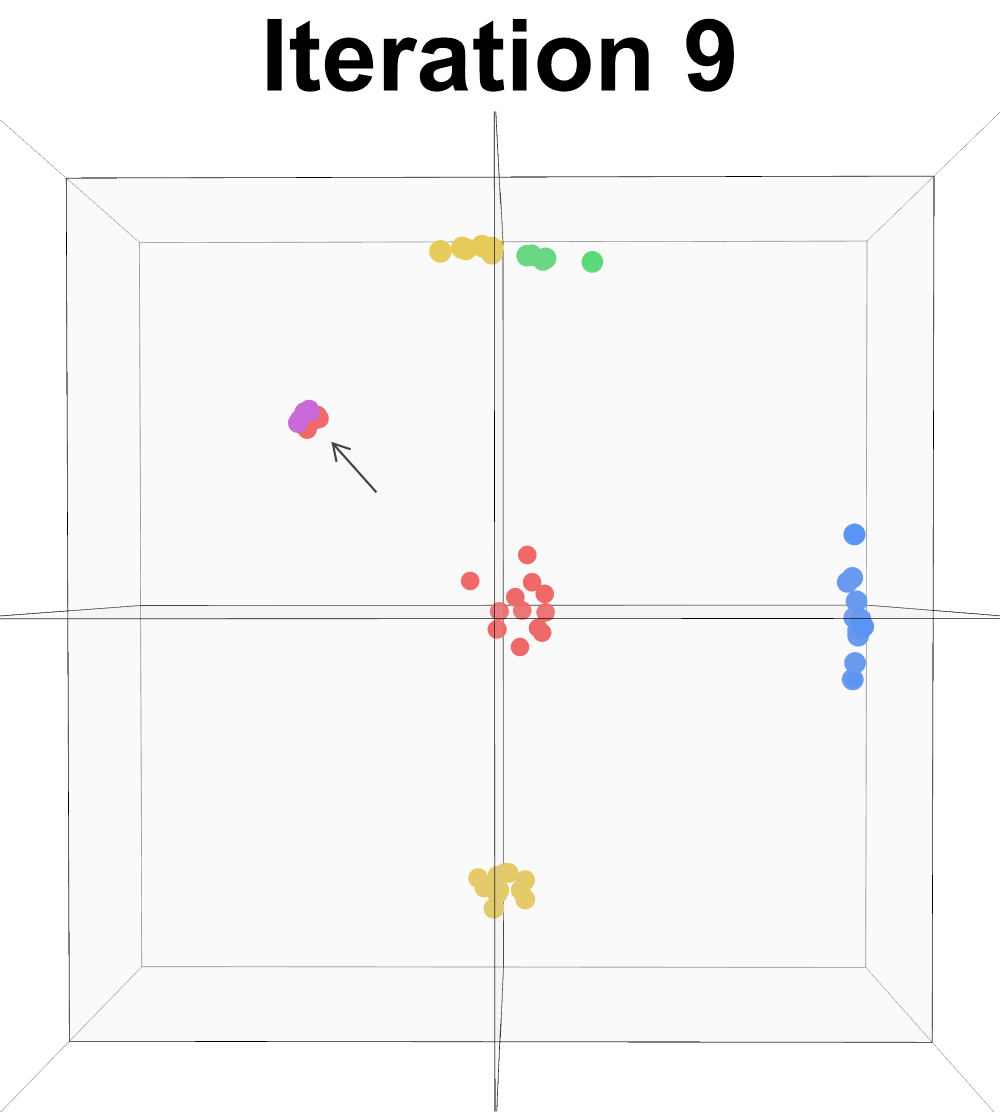}\\[0.5em]
        \includegraphics[width=\linewidth, trim={0 246 0 0cm}, clip]{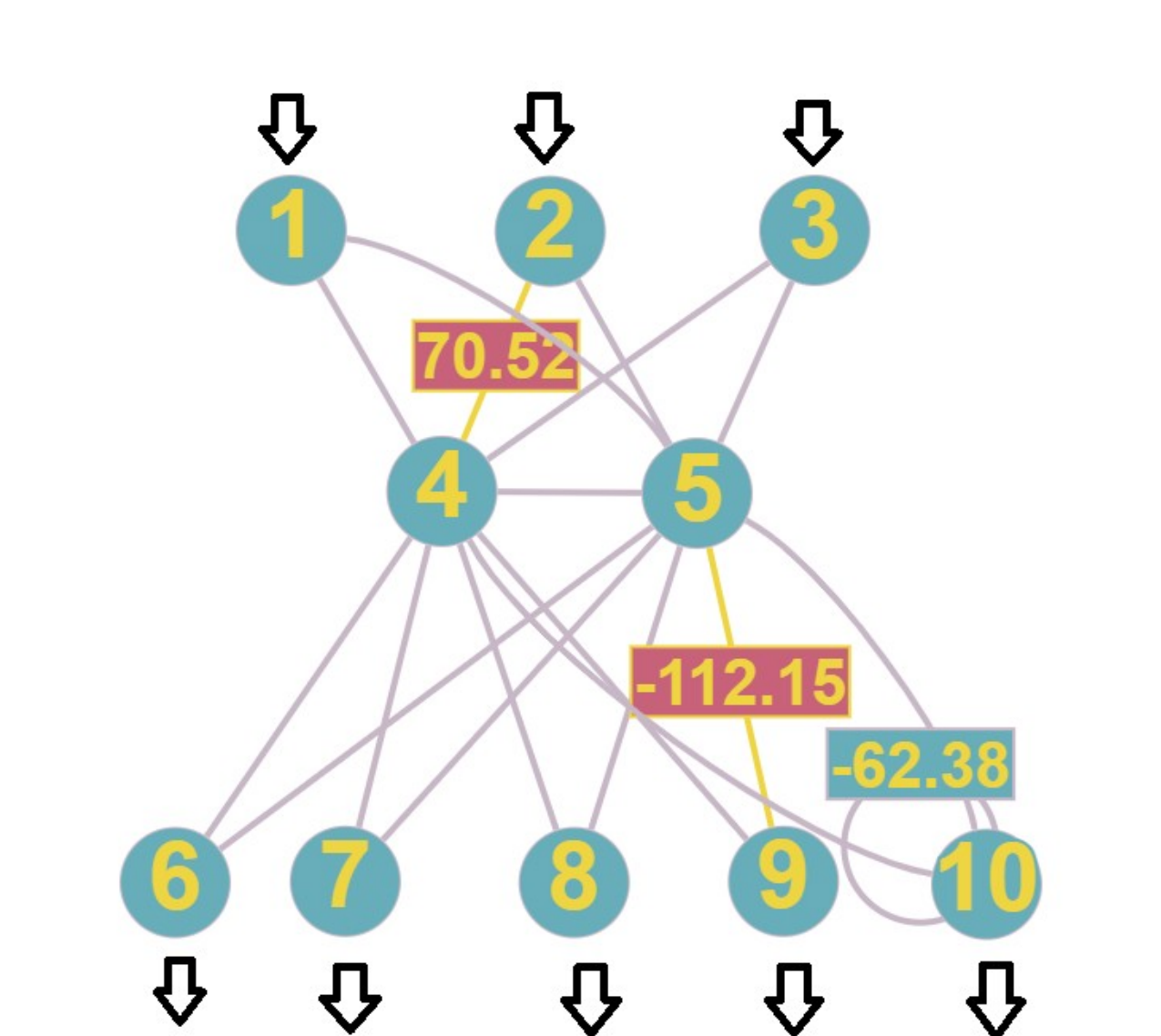}
        \label{fig:col-b}
    \end{minipage}
    \hfill
    \begin{minipage}[t]{0.3\linewidth}
        \centering
        \textbf{(c)} \\[0.3em]
        \includegraphics[width=\linewidth]{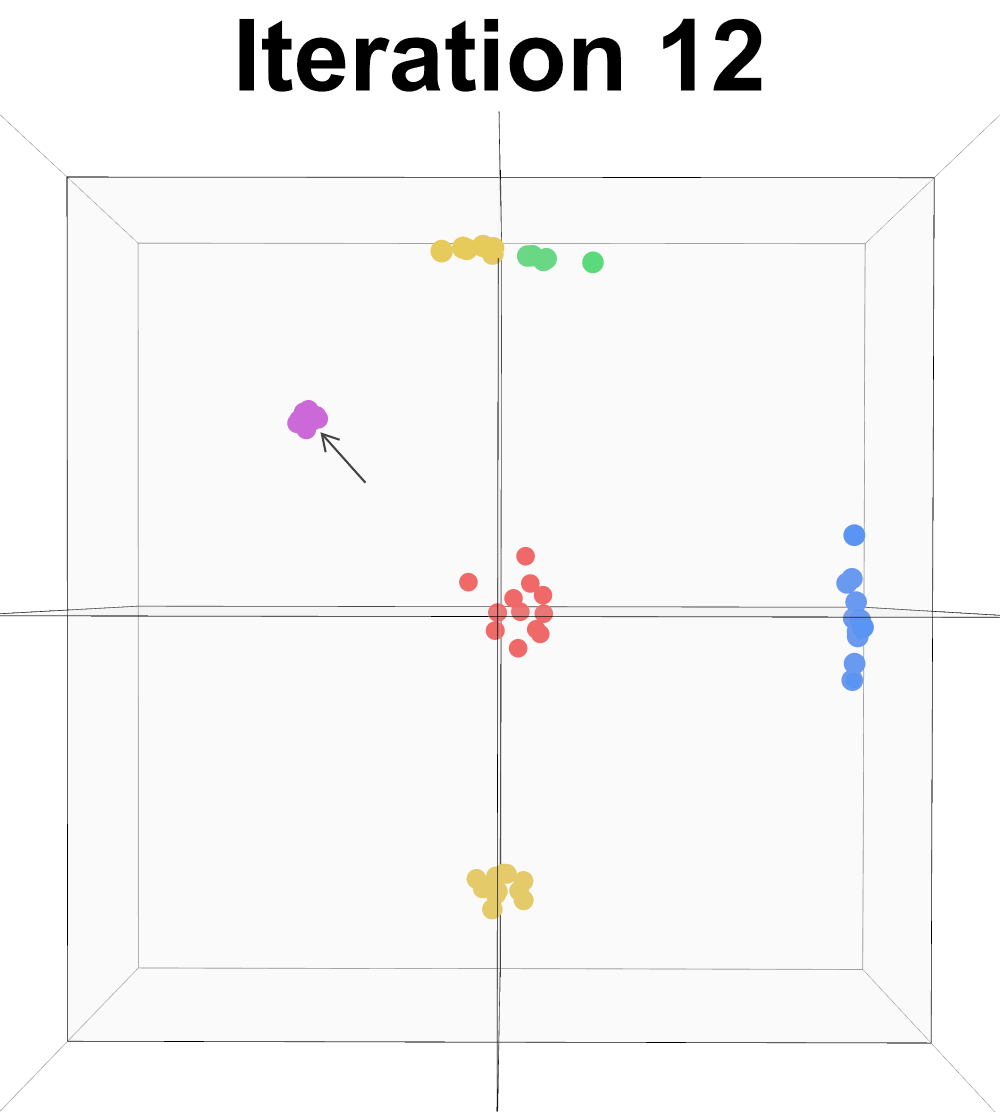}\\[0.5em]
        \includegraphics[width=\linewidth, trim={0 246 0 0cm}, clip]{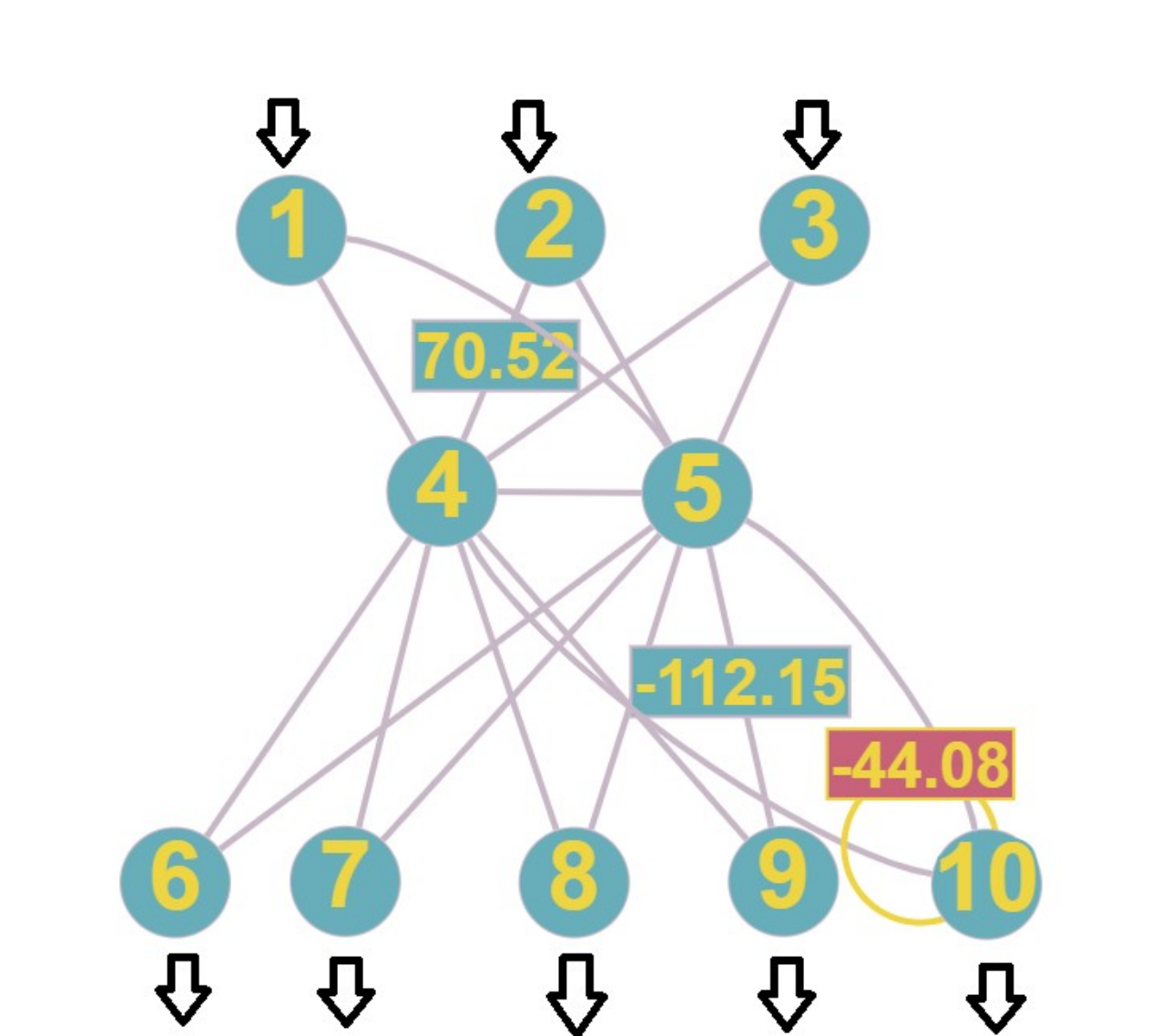}
        \label{fig:col-c}
    \end{minipage}
    
\caption{Frames captured from the Qlustering process applied to 60 state vectors in a 3-dimensional Hilbert space, grouped into 5 distinct clusters. Upper panels illustrate the spatial distribution of the vectors in the space, while the bottom panels illustrate the Qlustering network with its 3-2-5 node structure. The fitting process runs over iterations $t = 1, \dots, T$, where $T$ is a predefined number of steps. Panel (a) shows the initial state at $t = 1$, where 3 clusters are already correctly grouped. The corresponding Hamiltonian for this initial step is displayed in bottom panel of (a). In each subsequent iteration, a random entry in the Hamiltonian $\mathcal{H}$ is selected and modified. If the new Hamiltonian yields a lower cost function $CF$ (Eq.~\ref{CF}), it is retained and used in the next iteration. Bottom panel (b) presents the Hamiltonian at $t = 9$, with the two changes from the initial Hamiltonian highlighted in yellow. The resulting clustering at $t = 9$ is shown in upper panel (b). The panels of (c) illustrate the clustering and Hamiltonian at iteration $t = 12$. For the full clustering video, see Supplementary Movie 1.
}
    \label{figure 1}
\end{figure}

\end{widetext}

To illustrate the method (see Fig.~\ref{figure 1}), consider a set of 60 states in an $L=3$ Hilbert space distributed into $q=5$ clusters. The algorithm is provided with the number of clusters $q$ and the state vectors $\Psi = \Psi_1, \Psi_2, \dots, \Psi_N$. An initial random Hamiltonian - depicted graphically in Fig.~\ref{figure 1}(a) bottom - is used to represent the network and compute the initial steady-state currents. This defines the first clustering, as shown in Fig.~\ref{figure 1}(a) top, where each point represents a state vector’s position in Hilbert space and its color denotes the assigned cluster. In the next iterations, a single element of $\cH$ is modified, the (dynamical) cost function is evaluated, and the new Hamiltonian is accepted if the cost was reduced; the updated clustering after 9 iterations (along with the changes in the Hamiltonian, marked in red) is shown in Fig.~\ref{figure 1}(b). Fig.~\ref{figure 1}(c) displays the clustering and Hamiltonian after 12 iterations. The process continues until convergence or a predefined number of iterations is reached. A full video of the Qlustering process is provided as Supplementary Movie 1.

{\bf The quantum network.--} As mentioned, we consider a one-particle tight-binding Hamiltonian of the form 
\begin{equation}
\label{eq: Hamiltonian}
    \mathcal{H} = \sum_{i,j} h_{ij} c^\dagger_i c_j + \text{h.c.}
\end{equation} This general framework can model a variety of quantum transport networks, including (but not limited to) electron transport in quantum dots \cite{sarkar2022emergence}, exciton transport in biological systems \cite{zerah2021photosynthetic}, photon propagation in waveguides \cite{mookherjea2001optical,chen2021tight},   programmable nanophotonic processors \cite{harris2017quantum} and qubit networks \cite{weiss2021variational}.\\

To model propagation through the network, we use the Gorini-Kossakowski-Sudarshan-Lindblad (GKSL) quantum master equation \cite{lindblad1976generators,gorini1976completely}:
\begin{eqnarray}
\label{Lindblad}
\dot \rho &=& -i[H, \rho] + \sum_k \left( V_k^{\dagger} \rho V_k - \frac{1}{2}(V_k^{\dagger} V_k \rho + \rho V_k^{\dagger} V_k) \right) \\ \nonumber
&=& -i[H, \rho] + \mathcal{L}[\rho]
\end{eqnarray}
Here, $\rho$ denotes the system's density matrix, and $V_k$ are the Lindblad operators. In this setup, we define two Lindblad operators. The input operator is given by $V_{\text{in}} = \gamma_{\text{in}}^{1/2} \sum_{i=1}^{L} \Psi(i) c^\dagger_i$ \cite{zerah2021photosynthetic}, where $c^\dagger$ is the creation operator, $\gamma_{\text{in}}$ is the input dissipation rate, and $\Psi(i)$ is the $i$-th component of the input state vector $\Psi$. This operator represents a superposition of $\Psi$ injected into the input nodes.

The output operator is defined as $V_{\text{out},n} = \gamma_{\text{out}}^{1/2} c_r$, where $\gamma_{\text{out}}$ is the output dissipation rate and $c_r$ is the annihilation operator at the output node $r$. See Fig.~\ref{figure 2} for a schematic illustration.

The steady-state current resulting from this setup is computed as detailed in the Methods Section~\ref{METHODS}. This current serves as the input to the cost function defined in Eq.~\ref{CF}. After tuning the network parameters and clustering the dataset, the next step involves evaluating the structure of the data and validating the algorithm’s performance.

\textbf{Data structure and validation.--} To assess the performance of the Qlustering algorithm, it is first necessary to define the nature and structure of the data being clustered. In this study, following Fahad et al.~\cite{fahad2014survey}, we examine three types of data: synthetic data with tunable convexity, physical data with relatively high dimensionality (10 parameters per data point), and two examples of real-world datasets - QM9 \cite{ramakrishnan2014quantum} and Iris \cite{uci_iris}.\\

To evaluate clustering performance, we employ external validation metrics such as the Rand Index (RI) and Adjusted Rand Index (ARI)~\cite{sahlgren2005introduction,hubert_arabie_1985}, as well as internal validation metrics including compactness (CP) \cite{macqueen_1967,fahad2014survey}, the Dunn Validity Index (DVI) \cite{bezdek1998some,dunn_1974}, silhouette score \cite{batool2021clustering,rousseeuw_1987}, and clustering stability measured via label alignment using the Hungarian algorithm \cite{liu2022stability,fahad2014survey}. Descriptions and formulas for each metric are provided in Sec. ~\ref{METHODS}.

To improve robustness when repeated runs yielded inconsistent assignments, we applied an additional consensus-clustering post-processing protocol. This involved running the algorithm ten times on identical data and computing both the mean RI and ARI. In addition, we used hierarchical clustering average linkage to assign final groups from the consensus matrix following the technique of Monti et. al. \cite{monti2003consensus} (see Fig.~\ref{fig1}b and Fig.~\ref{fig:OverlapBarrier}b). Further details on this protocol are provided in Sec.~\ref{METHODS}.

\subsection{Classical--quantum workflow}
Although the clustering mechanism is determined by quantum transport through the network, the overall Qlustering procedure operates as a hybrid classical--quantum loop. At a broad level, the clustering dynamics themselves, namely the flow of information through the network and the resulting output currents, are produced by the quantum system, whereas data preparation, training, and evaluation are carried out classically. More specifically, the input data are first classically preprocessed and normalized into state vectors $\Psi_n$, after which each state is injected into the network and propagated according to the GKSL dynamics until the steady-state output currents are obtained. These currents constitute the quantum-computed observables used for cluster assignment, while the evaluation of the cost function, the acceptance or rejection of parameter updates, and the convergence checks are performed classically. In this sense, the quantum network serves as the physical computational core, whereas a classical controller carries out training, bookkeeping, and the optimization of the Hamiltonian entries $h_{ij}$. This division is important from the implementation perspective, since it requires only classical control over state preparation, readout, and network reconfiguration. At the same time, not all parts of this hybrid structure play the same role: state preparation, transport, and current readout are central to the method itself, while the precise form of the training protocol, preprocessing pipeline, and post-processing steps may be modified or improved without changing the underlying principle of Qlustering. In particular, some elements, such as the cost-function evaluation, are naturally classical, whereas others, especially the training strategy and the classical--quantum interface surrounding it, may admit more efficient hybrid realizations in future implementations. Post-processing procedures such as consensus clustering and validation-metric evaluation likewise remain fully classical.

\section{RESULTS}
\label{RESULTS}
\subsection{Clustering of points in 3D space}
\label{Position}

Our first example, though presented using state vectors, demonstrates a generalizable case applicable to diverse
data types. The inputs, $\Psi$, are vectors of length $d$, normalized such that their norm is unity, i.e. all points lie on the $(d-1)$-dimensional unit sphere. For demonstration, the $N$ input vectors are constructed as follows. We define $q$ ``base points'' $\{b_i\},\, i=1,\ldots,q$, and partition the $N$ data points into $q$ equal groups. In each group, vectors are drawn from a distribution centered on one of the base points with width parameter $0<\omega<1$. Each point is generated as $\Psi_n = (1-\omega)\,b_i + \omega\,u_n,$
where $u_n$ is a random unit vector, followed by normalization onto the unit sphere. Clearly, if $\omega$ is much smaller than the typical distance between base points, the groups remain well separated; for sufficiently large $\omega$, the groups overlap. At $\omega=1$, points are uniformly distributed on the unit sphere.

We evaluate Qlustering on three-dimensional synthetic data with $q = 4$, defined by the base points
$b_i$, $i=1,\ldots,4$, given by
$
b_1 = \{0.99,0.11,0.11\},\quad
b_2 = \{0.11,0.99,0.11\},\quad
b_3 = \{0.11,0.11,0.99\},\quad
b_4 = \tfrac{1}{\sqrt{3}}\{1,1,1\}.
$
Effective inter-cluster overlap begins at $\omega \simeq 0.3$. Perfect clustering was obtained for small $\omega$, decreasing to $\mathrm{RI} = 0.85$ and $\mathrm{ARI} = 0.63$ at $\omega = 0.3$.

\begin{widetext}
    
\begin{figure}
    \centering
    \includegraphics[width=0.9\linewidth]{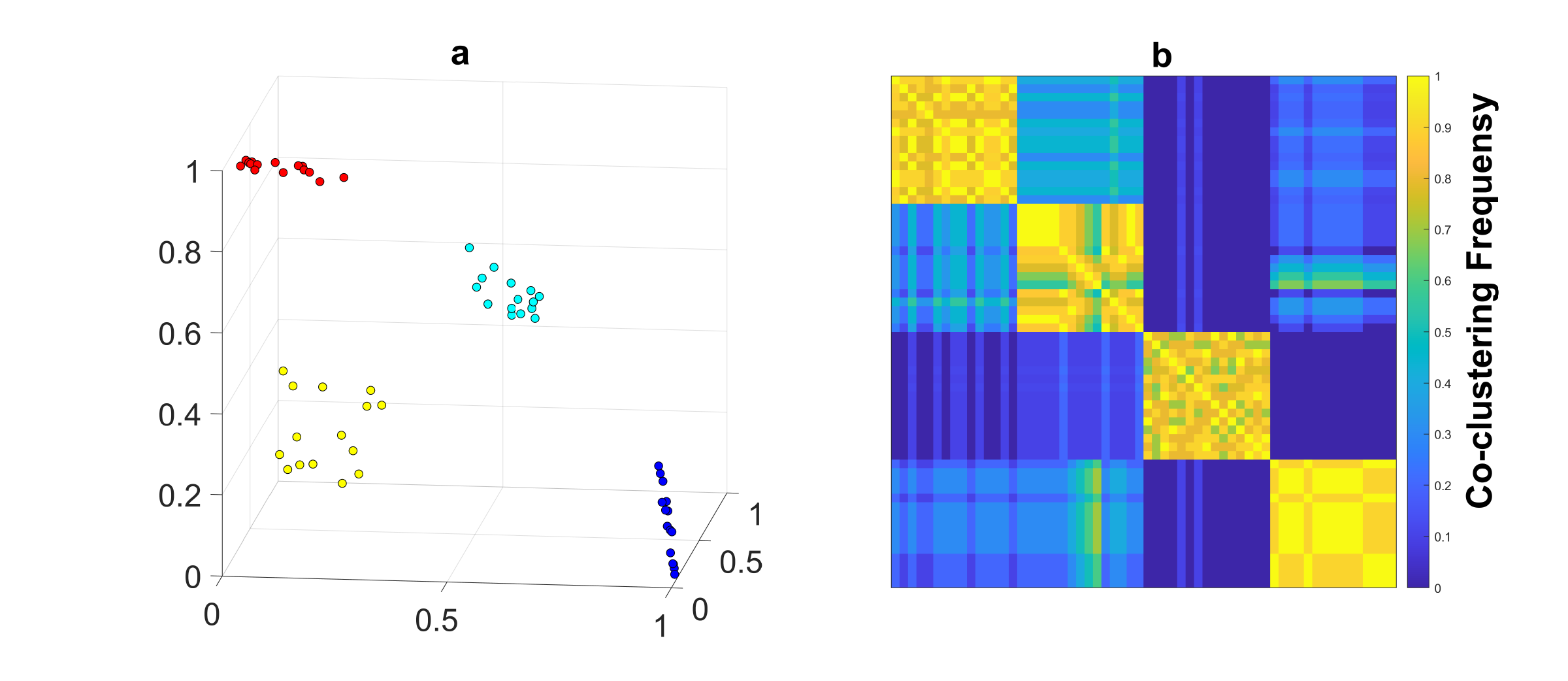}
\caption{\textbf{Qlustering of 3-dimensional vectors into four groups at $\omega = 0.15$.} 
(a) Spatial distribution of the input state vectors. (b) Consensus matrix from 10 repeated Qlustering runs. Yellow regions indicate high co-clustering frequency, while blue denotes low agreement. Each square represents a vector pair, with color intensity indicating the frequency of co-clustering (yellow: high consistency; blue: low agreement). A stable Qlustering pattern is observed, with four distinct block structures emerging in the consensus matrix, reflecting strong consistency in group assignments across runs.
} 
    \label{fig1}
\end{figure}
\end{widetext}

In Fig.~\ref{fig1}(a) we plot the spatial distribution of the 60 input state vectors, color-coded to reflect the predetermined classification (which is visually clear for $\omega = 0.15$, since the groups are well-separated). For these data, $RI=ARI=1$, i.e. essentially perfect clustering. Fig.~\ref{fig1}(b) shows the consensus matrix from 10 repeated Qlustering runs. Yellow regions indicate high co-clustering frequency, while
blue denotes low agreement. Each square represents a vector pair, with color intensity indicating the frequency of co-clustering
(yellow: high consistency; blue: low agreement). A stable Qlustering pattern is observed, with four distinct block structures emerging in the consensus matrix, reflecting strong consistency in group assignments across runs. 

To further challenge the algorithm and demonstrate Qlustering's capabilities with a larger number of classes in higher dimensions, we evaluated it on five groups ($q=5$) in a three-dimensional Hilbert space. Sixty data points were centered at $\nya{0,1,0}$, $\nya{0,0,1}$, $\nya{1,0,0}$, $\tfrac{2 \sqrt{3}}{9}\nya{-1.5,1.5,1.5}$, $\tfrac{2}{\sqrt{13}}\nya{0,1,1.5}$
, with varying group width $0<\omega<1$. 

\begin{figure}[t]
    \centering
    \includegraphics[width=\linewidth]{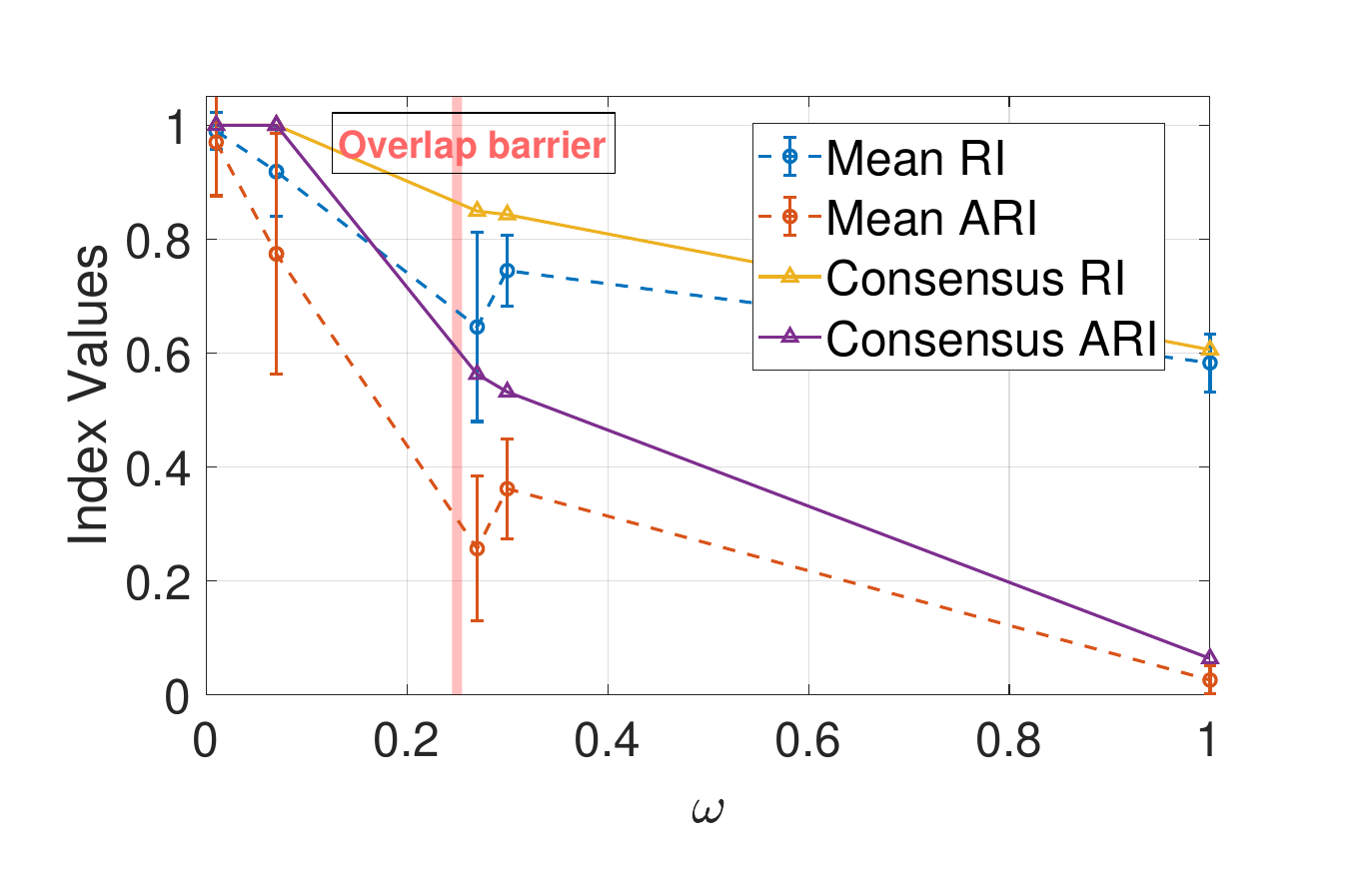}\\[0.5em] 
    \includegraphics[width=\linewidth]{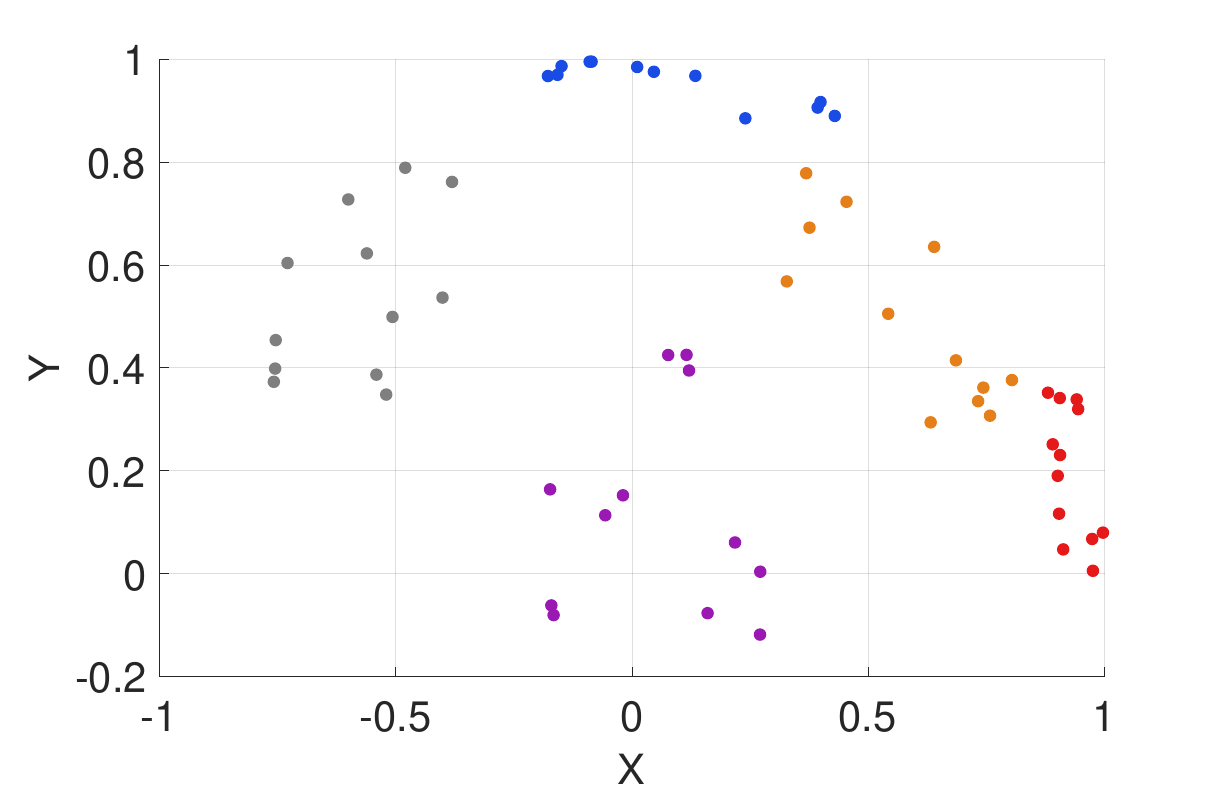}       
    \caption{\textbf{Top: Qlustering scores across varying group widths, $\omega$.} Solid-line triangles indicate the consensus clustering scores for the Rand Index (RI) and Adjusted Rand Index (ARI), while dashed-line circles represent the corresponding mean values. High clustering performance is maintained until significant overlap between groups occurs (marked by the transparent red line), after which scores decline toward random clustering levels as the distribution approaches uniformity. \textbf{Bottom: The overlap point visualized.} A 2D projection of the spatial distribution of five groups in a three-dimensional Hilbert space at $\omega = 0.25$, marking the onset of the overlap point. Colors indicate the clusters determined by the Qlustering algorithm. The groups remain largely separated, i.e. no true overlap yet, though some points already lie at approximately equal distances from their own center and that of another group. This regime marks the transition where Qlustering performance begins to decline from perfect scores toward reduced accuracy, as shown in the top figure.}
    \label{fig:OverlapBarrier}
\end{figure}

In Fig.~\ref{fig:OverlapBarrier} we plot the different clustering index values (RI and ARI, mean and consensus) for different distribution widths $\omega$ (top panel). For small widths, Qlustering achieved perfect clustering ($\mathrm{RI} = \mathrm{ARI} = 1$), while increasing the width reduced performance. Using the consensus scheme, perfect scores persisted over a wider range of widths, with performance declining only once substantial overlap occurred (so called overlap barrier). The bottom panel of Fig.~\ref{fig:OverlapBarrier} shows a 2D projection of the input vectors at $\omega=0.25$, marking the onset of inter-cluster overlap. Different colors indicate different clusters (predetermined by their initial base point, $b_i$), and one can recognize that the red and orange clusters will readily mix once $\omega$ is further increased. 

\subsection{Localization}
\label{Localization}

Overlap clustering is based essentially a geometric property, namely the distance (in Hilbert space) between different vectors. To demonstrate the versatility of the Qlustering algorithm to go beyond geometrical problems, and its ability to perform on physical, high-dimensional data, we applied it to the localization problem described  in Ref.~\onlinecite{lorber2024using}. In localization, we refer to whether the weight of the vector (normalized to one) is concentrated on one or a few entries, or instead distributed evenly across the entire vector. A key measure for the degree of localization is the Inverse Participation Ratio (IPR), \cite{kramer1993localization}
\begin{equation}
    IPR[\Psi_n]=\left(\sum^{L}_{i=1}|\Psi_n(i)|^4\right)^{-1}
\end{equation} 
which ranges between 1 and $L$ and quantifies the spatial extent of a quantum state, with higher values indicating strong delocalization (or an "extended state") and lower values corresponding to localized states. In the extreme limits, a uniformly spread vector will have $IPR=L$, and a fully localized state (one entry is 1 and the rest are 0) will have $IPR=1$. 

To adapt Qlustering to this task, we modified the cost function by defining 

\begin{equation*}
J_{\mathrm{tags}} =
\begin{cases}
[1,0], & J_1 \in [0.4, 0.6] \ \text{and} \ J_2 \notin [0.4, 0.6] \\
[0,1], & J_2 \in [0.4, 0.6] \ \text{and} \ J_1 \notin [0.4, 0.6] \\
[0.5,0.5], & \text{otherwise}.
\end{cases}
\end{equation*} and setting 
\begin{equation}
CF(J, N) = \sum_{n=1}^N (J_{\text{tags}} - J_n)^2 
\label{IPR cost}
\end{equation}

This formulation assigns extreme current values to one class and moderate (ambiguous) values to the other, effectively separating strongly localized states from delocalized ones. 

In the following examples, we generate 50 random vectors with $d=10$, and we wish to examine whether Qlustering can differentiate between extended and localized states. To make an artificial separation between the clusters, we select on vectors with $IPR$ lower than some value $\xi$ or higher than some value $L-\xi$. We define $\Delta_{\mathrm{IPR}} = IPR^{\mathrm{lowest}}_2 - IPR^{\mathrm{highest}}_1$, where $IPR^{\mathrm{highest}}_1$ and $IPR^{\mathrm{lowest}}_2$ denote the most extreme IPR values in each group (i.e. the highest IPR in the localized group and the lowest IPR in the extended group). The algorithm attains high accuracy when $\Delta_{\mathrm{IPR}}$ is large and the classes are well separated. As the overlap between groups increases, $\Delta_{\mathrm{IPR}}$ diminishes, leading to a decline in classification accuracy.

\begin{figure}[t]
    \centering
    \includegraphics[width=\linewidth]{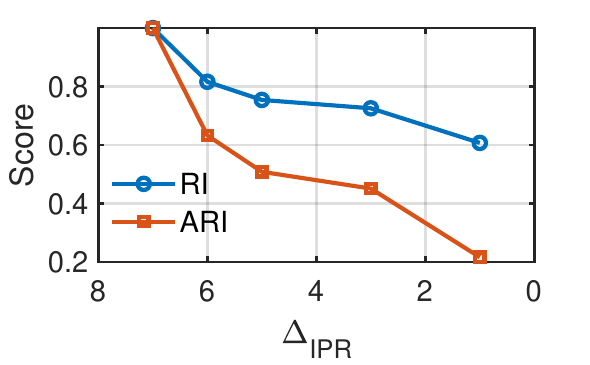}
    \caption{\textbf{Qlustering performance on the localization task.} Rand Index (RI) is shown in blue and Adjusted Rand Index (ARI) in orange on varying $\Delta_{IPR}$. The network clusters 10-dimensional state vectors with IPR values ranging from 1 to 10. As $\Delta_{IPR}$ decreases, performance declines. This drop corresponds to the mixing of strongly localized and delocalized states, which reduces the distinctiveness of the underlying physical classes.}

    \label{fig IPR}
\end{figure}

In Fig.~\ref{fig IPR} we show the RI and ARI as function of class separability $\Delta_{\mathrm{IPR}}$, for Qlustering of 50 ten-dimensional random vectors (taken from a uniform distribution) for $\Delta_{\mathrm{IPR}}=1,3,5,6,7$. As expected, for very high values of $\Delta_{\mathrm{IPR}}$ an essentially perfect clustering is achieved ($RI=ARI=1$). It then rapidly falls off as $\Delta_{\mathrm{IPR}}$ is reduced, yet in a non-linear way, reaching $0.7<RI<0.8$ even when $\Delta_{\mathrm{IPR}}$ is such that vectors from the localized and extended groups overlap substantially.

\subsection{QM9 dataset}
\label{QM9}
To evaluate Qlustering on complex, high-dimensional chemical data, we used the QM9 benchmark dataset, which contains more than 130000 small molecules with quantum‐chemical annotations \cite{ramakrishnan2014quantum,ruddigkeit2012enumeration}. Each entry includes a SMILES representation \cite{weininger1988smiles}, HOMO–LUMO energy gaps, temperature‐dependent internal energies, and other descriptors. Our objectives were: (i) to assess Qlustering’s robustness on realistic chemical informatics data, and (ii) to identify descriptor parameters that co-vary, potentially revealing structure–property relationships. Since Qlustering clusters points based on Hilbert‐space locality, the resulting groupings may reflect molecular features - such as symmetry or topological motifs - that influence specific chemical properties.

As input to the Qlustering algorithm, we randomly chose 97 molecules from the QM9 dataset, which were encoded using their Sorted Interatomic Distances (SID) \cite{grigoryan2003structure,sadeghi2013metrics}, computed from the 3D coordinates $\mathbf{r}_i \in \mathbb{R}^3$ of all non-hydrogen atoms:
\[
\mathrm{SID} = \mathrm{sort}\Bigl(\bigl\{ \lVert \mathbf{r}_i - \mathbf{r}_j \rVert_2 
\;\big|\; i < j,\ Z_i, Z_j \neq 1 \bigr\}\Bigr),
\]
where $Z_i$ denotes the atomic number of atom $i$. This permutation‐invariant descriptor compactly encodes molecular geometry and has shown strong performance in small‐molecule regression tasks \cite{Hansen2015}. Each molecule was thus represented by a fingerprint vector of length~10, corresponding to the maximum number of non-hydrogen interatomic pairs observed among the 97 molecules selected from the dataset. Shorter vectors were zero-padded to this length to ensure uniform dimensionality.


The molecular fingerprint vectors were normalized into state vectors $\Psi$ satisfying $\sum_j |\Psi(j)|^2 = 1$. Clustering results were evaluated using external metrics (RI, ARI) and internal metrics (compactness, Dunn Index, silhouette, stability), following Sec.~\ref{METHODS}. 

\begin{figure}[!t]
\centering
\includegraphics[width=\linewidth]{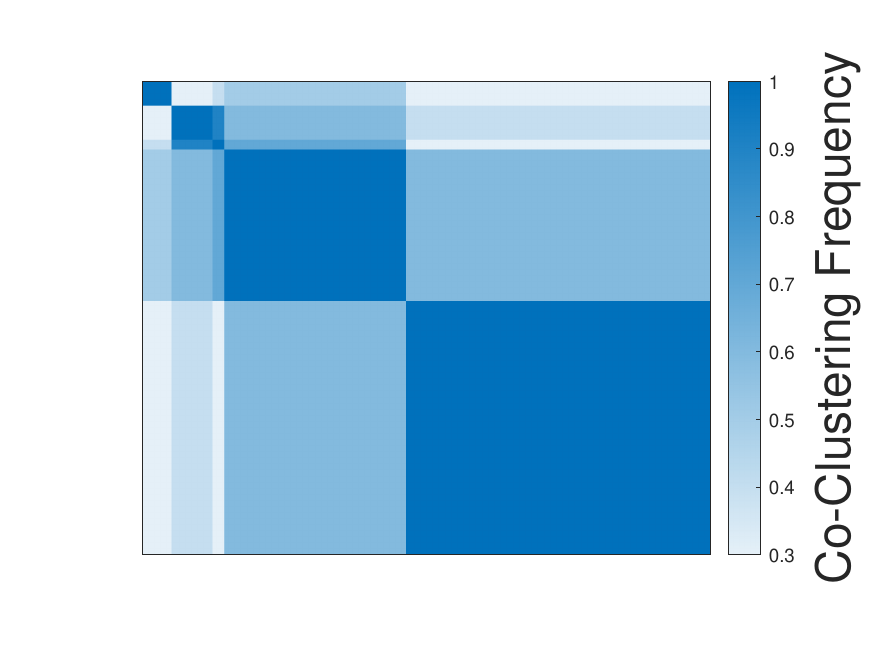}
\caption{\textbf{Consensus heatmap for Qlustering on a QM9 subset with $q=2$.} Pairwise consensus values from 10 consecutive Qlustering runs on a subset of 97 QM9 molecules are shown. Darker blue indicates a higher tendency for molecule pairs to cluster together, while lighter blue indicates lower co-clustering frequency. The network architecture (10--3--2) outputs two clusters; however, the consensus heatmap reveals a latent four-cluster structure. Running Qlustering with $q = 4$ confirmed this pattern and yielded improved internal clustering metrics, as discussed in the main text.}
\label{fig:QM9_consensus}
\end{figure}

Qlustering was first performed with $q=2$ groups, reaching internal scores of
 compactness = 19.87, Dunn Index = 0.765,
silhouette = 0.684. Consensus clustering based on the RotC descriptor resulted in external scores of $\mathrm{RI} = 0.75$ and $\mathrm{ARI} = 0.30$, while an individual run achieved up to $\mathrm{RI} = 0.90$ and $\mathrm{ARI} = 0.76$ using the same descriptor. In Fig.~\ref{fig:QM9_consensus} we plot the consensus heatmap of 10 consecutive Qlustering runs, performed on the 97 QM9 molecules described above, clustered into two groups ($q=2$). From the consensus matrix emerge four separate consensus subspaces, indicating that probably Qlustering into $q=4$ groups will outperform $q=2$ Qlustering.

\begin{figure}[!t]
\centering
\includegraphics[width=1\linewidth]{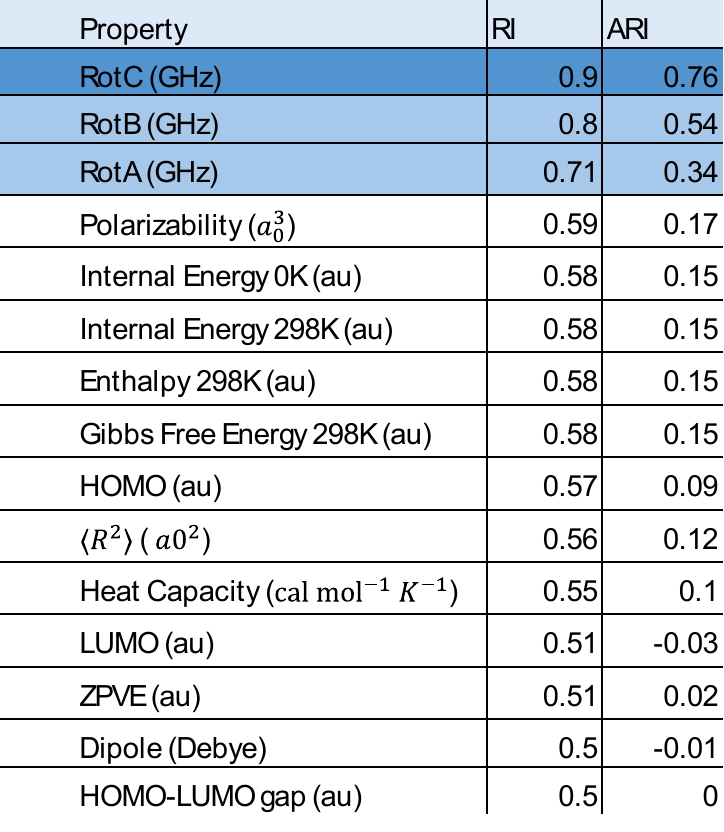}
\caption{\textbf{External clustering scores for a single Qlustering run for 14 molecular properties from the QM9 dataset ($q=2$).}  
Rand Index (RI) and Adjusted Rand Index (ARI) values are shown for each binarized property, computed on a subset of 97 molecules. Binarization was performed by tagging each input vector as above or below the mean value of the corresponding descriptor. Rotational constants A, B, and C yield the highest agreement with known labels, with RotC achieving $\mathrm{RI} = 0.90$ and $\mathrm{ARI} = 0.76$. These properties depend directly on molecular geometry, consistent with Qlustering’s structural input.}
\label{fig QM9 properties}
\end{figure}

External evaluation scores \textit{from a single run} for 14 binarized molecular descriptors in the $q = 2$ case are presented in Fig.~\ref{fig QM9 properties}. The RI and ARI (Eqs.~(\ref{eq: RI}), (\ref{eq: ARI})) were computed for each relevant molecular descriptor in the QM9 dataset: for every descriptor, its mean value was calculated, and input vectors were tagged as either above or below the mean. This procedure was repeated independently for all descriptors.
  The descriptors were then ranked according to these scores to identify the output variable that best aligns with the network’s classification. As the input vectors were derived from molecular structures, we expected the most predictive descriptor to exhibit a strong correlation with the underlying atomic configuration. The highest agreement was obtained for the rotational constant C (RotC), with best $\mathrm{RI} = 0.90$ and $\mathrm{ARI} = 0.76$, while rotational constants A and B also scored highly ($\mathrm{RI} > 0.70$, $\mathrm{ARI} > 0.30$). Rotational constants $(A, B, C)$, given by  
$A = \frac{h}{8\pi^2 c I_a}$,  
$B = \frac{h}{8\pi^2 c I_b}$,  
$C = \frac{h}{8\pi^2 c I_c}$,  
are inversely proportional to the principal moments of inertia $I_{a,b,c}$, where $h$ is Planck’s constant and $c$ is the speed of light. As these moments depend on the spatial distribution of atomic masses, the constants provide a direct probe of molecular geometry. This aligns with our aim of using Qlustering as a covariance detector, identifying correlations between structural features and physical properties.
Stability across runs reached $0.754$, exceeding typical benchmarks \cite{fahad2014survey}. 

Based on the consensus heatmap (Fig.~\ref{fig:QM9_consensus}), which showed 4 stable clusters, we tested Qlustering also with $q=4$. For each case, 10 independent Qlustering runs were conducted, following the procedure in Sec.~\ref{Position}. Internal scores for $q = 4$ were  compactness $= 0.865$, Dunn Index $= 2.4251$, silhouette $= 0.9776$, and stability $= 0.80$. These internal scores are better than for $q=2$, and a comparison to k-means (Table \ref{tab:kmeans_scores}) shows a distinct advantage for Qlustering.

\subsection{Iris dataset}
\label{iris}
The Iris dataset, introduced by Ronald A.~Fisher in 1936 \cite{fisher1936iris} and obtained from the UCI Machine Learning Repository~\cite{uci_iris}, is a classic benchmark in pattern recognition and unsupervised learning. It contains 150 samples from three Iris species (\textit{Iris setosa}, \textit{Iris versicolor}, and \textit{Iris virginica}), each described by four features: sepal length, sepal width, petal length, and petal width. Although the dataset includes labeled classes, it is widely used in clustering tasks to assess an algorithm’s ability to recover natural groupings without prior knowledge. Its well-structured yet partially overlapping class distributions make it a useful testbed for comparing clustering methods \cite{gupta2018comparison, chakraborty2020comparative, mittal2019performance, ye2023comparison}.
The dataset was clustered using Qlustering, with internal and external evaluation scores measured over 10 consecutive runs. A network with a 4--2--3 architecture was used to learn feature representations conducive to clustering. Normalization transformed the input features into state vectors, which distorted the natural cluster structure and increased the difficulty of separation, as presented in Appendix  \ref{appendix: iris}. It is evident that normalization expands the spatial distribution of the groups, thereby reducing their separability. To address this, the sepal width feature was removed - a common practice when clustering the Iris dataset \cite{mani2024enhancing} - yielding a substantial improvement in consensus clustering performance, outperforming k-means “on its own field.”
Using all four features, Qlustering achieved $\mathrm{RI} = 0.77$, $\mathrm{ARI} = 0.56$, and internal metrics of stability $= 0.72$, compactness (CP) $= 458.3$, Dunn Validity Index (DVI) $= 0.038$, and silhouette score $= 0.14$ (all mean values over 10 runs). Removing the sepal width feature increased external scores to $\mathrm{RI} = 0.92$ and $\mathrm{ARI} = 0.82$, with internal metrics of stability $= 0.60$, CP $= 2.5$, DVI $= 0.022$, and silhouette score $= 0.38$.

The fact that normalization increases the separation difficulty highlights a potential weakness of Qlustering - the normalization step. However, with appropriate data preparation, this step can be mitigated and may even become an advantage for certain data structures.
\subsection{Assessment of k-means algorithm on the data}
\label{Classic}
The k-means algorithm, a well-established benchmark method, was applied to all the tasks described above to provide a comparative baseline for evaluating Qlustering’s performance and limitations. The standard \texttt{kmeans} function from MATLAB’s Statistics and Machine Learning Toolbox was used, with the corresponding code provided in the Supplementary Material.  Table~\ref{k-means} presents the internal and external evaluation metrics of the widely used k-means algorithm, applied to the same four clustering tasks explored with Qlustering in this study. As expected from a well-established and extensively tested method, k-means performs competitively in most cases. In fact, the two algorithms often yield comparable scores. However, several key differences in their behavior and strengths are worth noting:
\begin{itemize}[topsep=2pt, itemsep=2pt, parsep=0pt, leftmargin=10pt]
    \item \textbf{Compactness (CP):} K-means consistently outperforms Qlustering in terms of compactness. This is expected, as CP serves as the cost function optimized by k-means itself, giving it a natural advantage on this metric.
    
    \item \textbf{Clustering of Points in 3D space (Sec.~\ref{Position}):} In this synthetic setup, k-means achieves strong results even when $\omega = 0.3$, where the group boundaries begin to overlap. This suggests that Qlustering may struggle with convex or linearly separable data structures under certain configurations.
    
    \item \textbf{Localization problem (Sec.~\ref{Localization}):} In contrast, k-means performs poorly - almost randomly - on the localization task. This is consistent with known limitations of k-means in high-dimensional data spaces \cite{fahad2014survey}. Qlustering, in comparison, only shows degraded performance when the inter-group parameter range (IPR) gap is minimal.
    
    \item \textbf{QM9 dataset: (Sec.~\ref{QM9})} On this real-world dataset, k-means shows strong internal metrics but weak external agreement with known labels. This discrepancy further highlights the strength of Qlustering in uncovering meaningful parameter dependencies beyond geometric compactness.
    
    \item \textbf{Iris dataset (Sec.~\ref{iris}):} With all four features included, k-means outperforms Qlustering on most metrics. However, removing the sepal width feature - known to contribute to class overlap - leads to Qlustering surpassing k-means on the majority of external and internal scores.
\end{itemize}

Beyond the metric-by-metric comparison, an important practical advantage of Qlustering is its relatively high run-to-run stability. This feature also makes the final assignment more amenable to consensus clustering than methods that are more sensitive to initialization.

\begin{figure}
\centering
\includegraphics[width=\linewidth]{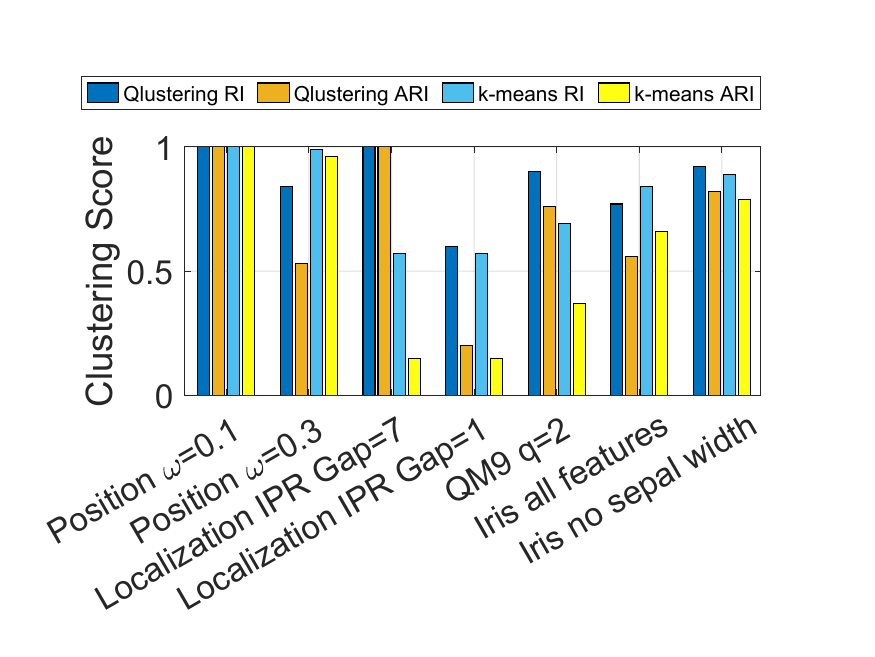}
\caption{\textbf{External clustering performance of Qlustering vs.\ k-means.}
Rand Index (RI) and Adjusted Rand Index (ARI) scores for both algorithms are shown across several datasets: positional clustering with $\omega = 0.1$ and $\omega = 0.3$, localization with $\Delta_{IPR}$ of 7 and 1, QM9 with $q = 2$, and the Iris dataset with and without sepal width. Qlustering performs competitively in most cases. Two notable exceptions are: (i) the positional task with $\omega = 0.3$, where k-means performs better, indicating difficulty in separating overlapping clusters; and (ii) the IPR task with 10 parameters, where Qlustering shows a clear advantage in high-dimensional, complex data. Further discussion appears in this Section and in Sec. ~\ref{summary}.} 

\label{fig:Qlustering_v_kmeans}
\end{figure}

A graphical illustration of the external comparison between Qlustering and k-means on the various datasets used here is shown in Fig.~\ref{fig:Qlustering_v_kmeans}. 

Our aim in this section is to demonstrate the competitive performance of Qlustering relative to currently used methods. Comparison with the well-established k-means algorithm serves this purpose effectively, providing a clear benchmark for evaluation. While testing additional clustering algorithms could further broaden the analysis and will be considered in future work, the main focus here is to show that Qlustering not only performs clustering but does so competitively against a standard and widely used reference method. For a broader qualitative comparison of clustering methods, see~\cite{fahad2014survey}, which employs the same metrics to evaluate various algorithms.

\begin{table*} 
    \centering
    \caption{K-Means Clustering Performance Scores}
    \label{tab:kmeans_scores}
    \begin{tabular}{|l|c|c|c|c|c|c|c|}
        \hline
        \textbf{Score} & \multicolumn{2}{c|}{\textbf{Position}} & \multicolumn{2}{c|}{\textbf{Localization}} & \multicolumn{2}{c|}{\textbf{QM9}} & \textbf{Iris} \\
        \cline{2-8}
        & \textbf{$\omega=0.1$} & \textbf{$\omega=0.3$} & \textbf{$\Delta_{IPR}=7$} & \textbf{$\Delta_{IPR}=1$} & \textbf{$q=2$} & \textbf{$q=4$} &  \\
        \hline
        \multicolumn{8}{|l|}{\textbf{External Scores}} \\
        \hline
RI & $1$ & $0.987$ & $0.571$ & $0.571$ & $0.693$ & $\,\_\_\_$ & $0.843$ \\
ARI & $1$ & $0.957$ & $0.150$ & $0.150$ & $0.373$ & $\,\_\_\_$ & $0.659$ \\
\hline
\multicolumn{8}{|l|}{\textbf{Internal Scores}} \\
\hline
CP & $0.482$ & $8.31$ & $14.109$ & $9.73$ & $15.434$ & $1.198$ & $91.905$ \\
DVI & $3.337$ & $0.132$ & $0.415$ & $0.257$ & $0.674$ & $2.188$ & $0.095$ \\
Silhouette & $0.981$ & $0.630$ & $0.417$ & $0.345$ & $0.697$ & $0.716$ & $0.716$ \\
\hline
Stability & $1.00$ & $0.790$ & $0.846$ & $0.750$ & $0.851$ & $0.940$ & $0.851$ \\

        \hline
    \end{tabular}
    \label{k-means}
\end{table*}
\subsection{Computational complexity}

It is useful to distinguish between the cost of classically simulating Qlustering and the cost of executing it on a physical device. In a classical simulation, the main bottleneck is the evaluation of the steady-state current for a given network. In the present formulation, this requires solving for the null space of a \(d^2 \times d^2\) Liouvillian matrix, where \(d\) is the total network dimension, i.e., the number of nodes. In a naive dense linear-algebra treatment, this scales as \(\mathcal{O}(n^3)\), which here gives \(\mathcal{O}((d^2)^3)=\mathcal{O}(d^6)\). This cost reflects the numerical overhead of simulating the open-system dynamics and should not be identified with the runtime of the physical protocol itself.

For a physical implementation, the relevant cost is instead set by input preparation, propagation to the readout regime, measurement of the output currents, and Hamiltonian reconfiguration during training. Let \(i\) be the number of training iterations, \(P\) the number of particles considered per iteration, \(N\) the number of input states, \(L\) the dimension of the input state vectors \(\Psi\), \(q\) the number of measured output ports, \(t_{\mathrm{st}}\) the relaxation time to steady state, and \(h\) the number of nonzero hopping terms in \(\mathcal{H}\), which scale with the network size \(d\). The relevant per-iteration contributions are:
\begin{itemize}
\item state preparation for \(N\) inputs \(\Psi_n\): \(\mathcal{O}(NL)\),
\item propagation to the readout regime: \(\mathcal{O}(N t_{\mathrm{st}})\),
\item measurement of the output currents: \(\mathcal{O}(Nq)\),
\item Hamiltonian reconfiguration: \(\mathcal{O}(h)\) for global reprogramming and \(\mathcal{O}(1)\) for a local update,
\item cost-function selection over the \(P\) particles: \(\mathcal{O}(P)\),
\item bookkeeping: \(\mathcal{O}(P)\).
\end{itemize}
Accordingly, the training cost per iteration is
\begin{equation*}
\begin{aligned}
T_{\mathrm{iter}}
&=
\mathcal{O}\!\big(P(NL + N t_{\mathrm{st}} + Nq + h + 1)\big) \\
&\simeq
\mathcal{O}\!\big(P(NL + N t_{\mathrm{st}} + Nq + h)\big).
\end{aligned}
\end{equation*}
and the total training complexity over \(i\) iterations is
\begin{equation}
\label{complexity}
\mathcal{T}_{\mathrm{train}}
=
\mathcal{O}\!\big(iP\,(NL + N t_{\mathrm{st}} + Nq + h)\big).
\end{equation}

For comparison, k-means, which counts as a very low-computational algorithm, is also linear in complexity with $\mathcal{O}(N \cdot k \cdot d \cdot i)$.

The main system-dependent contribution in Eq.~\eqref{complexity} is \(t_{\mathrm{st}}\). This quantity is not universal, but depends on the effective spectral gap \(\Delta\) of the Liouvillian associated with \(\mathcal{H}\) and on the coupling to the environment. As a result, \(t_{\mathrm{st}}\) may increase with network size and may become large when strict convergence to steady state is required. This contribution may be reduced, however, by suitable choice of physical platform or by identifying an optimized readout regime before full steady state is reached.

In the compact networks considered here, the number of intermediate nodes did not exceed five, so both \(d\) and \(h\) remained small. In addition, the cost function depends sensitively on the steady-state current \(J[\Psi_n]\), which reflects the eigenstructure of \(\mathcal{H}\). Consequently, small perturbations of \(\mathcal{H}\) can produce substantial changes in the cost function, and in practice the required number of training iterations \(i\) remained modest.

\subsection{Physical implementations}
\label{Physical_Implementation}
The results obtained here identify a concrete set of implementation requirements at the level of the effective model. Qlustering requires a platform that supports: tunable couplings defining the network Hamiltonian, preparation of input states at the injection ports, open-system injection and extraction, and readout of the terminal observables that determine the clustering outcome. In the present framework, these observables are the steady-state output currents. This is a practically relevant feature of the method, since it replaces full state tomography by local transport measurements and thereby removes the need to reconstruct the full density matrix or resolve detailed transient dynamics.

Among existing platforms, photonic transport architectures appear to provide the most direct correspondence to these requirements. In particular, integrated photonic transport experiments have already demonstrated random-walk dynamics closely related to those considered here \cite{caruso2016fast}, and programmable photonic processors provide tunable network connectivity together with controllable input preparation and terminal readout \cite{harris2018linear,harris2017quantum}. Within such platforms, the effective Hamiltonian is set by the programmable couplings, while the clustering output is obtained from the measured terminal intensities or fluxes associated with the transport pattern.

An additional implementation route is conceivable on general-purpose quantum hardware, including superconducting transmon-based platforms, although in a less native form. There, the required GKSL-type dynamics would need to be synthesized through state preparation, gate sequences, or engineered dissipation, together with measurement protocols adapted to steady-state observables. In this sense, the present results point more directly toward analog transport platforms than toward gate-based realizations.

The implementation picture is further supported by the dephasing analysis presented in Appendix~\ref{appendix: dephasing}. For the tasks considered in Sec.~\ref{Position} and Sec.~\ref{QM9}, Qlustering retains high RI and ARI values across a broad range of dephasing strengths. In particular, stable performance is maintained up to $\Gamma_{\mathrm{dep}} = 10^{4}$ in units of $\bar{t}^{-1}$, where $\bar{t}^{-1}$ denotes the characteristic hopping scale of the network [see Eq.~\ref{eq: Hamiltonian}]. These results indicate that the clustering protocol does not rely on a narrowly coherent regime and remains effective under open-system conditions relevant to realistic devices.

For reference, in superconducting transmon hardware the dominant energy scale is typically set by the Josephson energy $E_J$ and lies in the GHz range \cite{krantz2019quantum}. Taking $\bar{t}^{-1}$ to be of this order gives characteristic hopping times in the sub-nanosecond to nanosecond regime. While a direct mapping of the present model to a specific hardware design lies beyond the scope of this work, this scale comparison is consistent with the conclusion drawn from the dephasing analysis: Qlustering operates robustly in an open-system regime compatible with experimentally relevant energy and noise scales.

\section{DISCUSSION}
\label{summary}

In this work, we introduced Qlustering, an unsupervised learning scheme based on steady-state quantum transport in open quantum networks. The main result of this work is that unsupervised structure extraction can be implemented through steady-state open-system quantum transport. Rather than reconstructing a quantum state or tracking transient dynamics, Qlustering uses terminal currents as the experimentally accessible observables that define the clustering representation. This extends current-based quantum transport networks from supervised classification to label-free organization of data, showing that the same physical mechanism can support a broader class of learning tasks. We evaluated the method on four representative tasks: synthetic state vectors in Hilbert space, localization-based clustering using the inverse participation ratio (IPR), molecular data from the QM9 dataset, and the Iris dataset.

Performance was assessed using internal and external clustering metrics, together with an analysis of computational complexity. For comparison, k-means was applied to the same datasets. This allows us to summarize the behavior of Qlustering across several settings:

\begin{itemize}
    \item \textbf{Computational complexity---} The training cost of Qlustering is
    \[
    \mathcal{T}_{\text{train}}
    = \mathcal{O}\!\big(iP\,(N L + N t_{\mathrm{st}} + N q + h)\big),
    \]
    as given in Eq.~\eqref{complexity}, where $i$ is the number of training iterations, $P$ the number of particles considered per iteration, $N$ the number of input states, $L$ the dimension of the input vectors, $q$ the number of measured output ports, $t_{\mathrm{st}}$ the relaxation time to steady state, and $h$ the number of nonzero hopping terms in $\mathcal{H}$. The scaling is therefore linear in the dataset size $N$, similar in spirit to $k$-means, while the main system-dependent cost is set by $t_{\mathrm{st}}$, which depends on the Liouvillian spectral gap and may become large in slowly relaxing networks. In the compact networks studied here, however, both $h$ and the required number of iterations $i$ remained small, so the practical cost is governed mainly by the physical time needed to reach the readout regime.

    \item \textbf{Dataset scale---} Qlustering was tested here on relatively small datasets (up to 150 data points), primarily because classical simulation of the transport dynamics is expensive. The present results therefore do not yet establish performance at larger scales, although this limitation arises from simulation cost rather than from the conceptual structure of the method itself.

    \item \textbf{High dimensionality---} In Sec.~\ref{Localization}, we applied Qlustering to localization data with ten parameters per point. In this setting, the method remained effective and outperformed $k$-means on the corresponding benchmark.

    \item \textbf{Physical robustness---} A central feature of Qlustering is that computation is based on steady-state transport and does not require gate-based control during inference. This makes the framework naturally compatible with noisy open-system platforms and supports operation under dephasing, as examined in Appendix~\ref{appendix: dephasing}.

    \item \textbf{Type of datasets---} The main limitation observed here is sensitivity to overlap between groups. Across the tested examples, performance decreased as the underlying structure became less well separated, in some cases more markedly than for $k$-means. This suggests that the method is currently best suited to datasets with sufficiently distinct structure in the induced transport representation.

    \item \textbf{Internal validity—} Here we evaluated Qlustering using four internal metrics: compactness (CP), Dunn validity index (DVI), silhouette score, and stability. Across the tasks considered, the most distinctive internal advantage of Qlustering is its high stability across repeated runs, which remains relatively robust to Hamiltonian initialization and can be directly exploited through consensus clustering to improve the final assignment. The strongest internal performance is obtained for the QM9 dataset, where the consensus-indicated $q=4$ solution improves CP, DVI, silhouette, and stability relative to $q=2$, indicating that the transport dynamics resolve a finer latent structure than is visible in the original two-cluster setting. By contrast, internal validity is weaker for the Iris dataset, particularly when all four normalized features are retained, reflecting the distortion introduced by the normalization step (Appendix~A). In the synthetic benchmark, the internal scores also deteriorate as inter-group overlap increases. Taken together, these results show that Qlustering is most effective when the transport dynamics induce a sufficiently well-separated representation of the data, and that its strong run-to-run stability is a useful practical advantage of the method.

    \item \textbf{External validity---} External validation indicates that Qlustering performs consistently across most of the tasks considered here, with further improvement obtained through consensus clustering. In several cases, this consistency appears to benefit from reduced sensitivity to initialization compared with $k$-means, although this point merits further systematic study.
\end{itemize}

These results support the feasibility of Qlustering as a transport-based approach to clustering and to the analysis of parameter correlations. In the settings studied here, Qlustering exhibits robustness to dephasing, low computational complexity, and rapid convergence, and performs well on structured, relatively well-separated datasets, including high-dimensional examples.

From an implementation perspective, the present results suggest that Qlustering is most naturally suited to analog transport platforms that support programmable couplings, state injection, open-system extraction, and terminal readout. In such settings, the steady-state currents used for clustering can be accessed directly, avoiding the need for full state tomography and making the protocol closer to the native observables of the device. The dephasing analysis further supports this picture by showing that the method remains effective over a broad range of noise strengths, indicating that its operation does not rely on a narrowly coherent regime. Although realization on general-purpose quantum hardware is conceivable through synthesized GKSL-type dynamics, the correspondence is less direct, and the present results point more naturally toward open-system transport architectures such as programmable photonic networks.

Future work may address cluster-number estimation, broader classes of data and preprocessing strategies, and the optimization of network-specific hyperparameters.


\section{METHODS}
\label{METHODS}
\textbf{Current from the Lindblad equation}
We evaluate currents through a steady-state solution of the Lindblad equation. \cite{zerah2021photosynthetic, zerah2018universal, zerah2020effects, zerah2023signature, sarkar2020environment} We consider a system governed by a general tight-binding Hamiltonian, with dynamics described by the Lindblad master equation:
\begin{eqnarray}
\dot{\rho} = -i[\cH, \rho] + \sum_k \left( V_k^{\dagger} \rho V_k - \frac{1}{2} { V_k^{\dagger} V_k, \rho } \right) \nonumber \
= -i[\cH, \rho] + \mathcal{L}[\rho]~. \nonumber
\end{eqnarray}
The calculation is restricted to the single-exciton manifold, which suffices for capturing the relevant physics, as established in prior work \cite{zerah2021photosynthetic}.
 
We solve for the steady-state density matrix $\rho_s$, defined by $\dot{\rho}_s = 0$. The current $J_i$ at site $i$ is given by the continuity equation,
\begin{eqnarray}
J_i = \frac{d\langle n_i \rangle}{dt} = \frac{d}{dt} \mathrm{Tr}(\hat{n}_i \rho)~.
\end{eqnarray}
where $n_i$ is the number density in site $i$.
Taking the time derivative inside the trace yields:
\begin{eqnarray}
J_i = \mathrm{Tr}(\dot{\hat{n}}_i \rho_s + \hat{n}_i \dot{\rho}_s)~.
\label{eq:Ji}
\end{eqnarray}

At steady state the expression simplifies. For exit sites, where $\dot{\hat{n}}_i = -i[\hat{n}_i, \cH]$, the total current conservation implies:
\begin{eqnarray}
J_{ext} = \mathrm{Tr} \left( \hat{n}_{ext} \left( -i[\cH, \rho_s] + \mathcal{L}_{ext}[\rho_s] \right) \right) = 0~.
\label{eq:Jcons}
\end{eqnarray}

Due to current conservation, at steady state the current outside the network equals in magnitude to the current inside, and opposite in sign. Hence the zero net current. The outside current $J_{ext}$ can be evaluated solely from the term:
\begin{eqnarray}
J_{ext} = \mathrm{Tr} \left( \hat{n}_{ext} \mathcal{L}_{ext}[\rho_s] \right)~.
\label{eq:Jfinal}
\end{eqnarray}
\\
\textbf{External and internal validation metrics}

\paragraph{Rand index (RI) and Adjusted Rand index (ARI) scores} The RI measures the proportion of pairwise agreements,whether two points are correctly grouped or separated, and is defined as \cite{hubert_arabie_1985,sahlgren2005introduction}:
\begin{equation}
    \text{RI} = \frac{TP + TN}{TP + TN + FP + FN}
    \label{eq: RI}
\end{equation}

where \( TP \) and \( TN \) denote true positives and true negatives, and \( FP \), \( FN \) are false positives and false negatives. RI values range from 0 (complete disagreement) to 1 (perfect agreement).

However, RI does not account for agreement due to chance; high RI values may still occur in poorly performing models. To address this, we use the ARI \cite{hubert_arabie_1985}:
\begin{equation}
    \text{ARI} = \frac{\text{RI} - \mathbb{E}[\text{RI}]}{\max(\text{RI}) - \mathbb{E}[\text{RI}]}
    \label{eq: ARI}
\end{equation}

where \( \mathbb{E}[\text{RI}] \) is the expected RI of a random model. ARI ranges from \(-1\) to 1, with 0 indicating agreement by chance and 1 for a perfect match. Negative values mean the clustering is worse than random.
\paragraph{Compactness (CP)}  
Compactness measures the within-cluster dispersion by computing the sum of squared distances between data points and their respective cluster centroids \cite{macqueen_1967,fahad2014survey}:
\[
\text{CP} = \sum_{i=1}^{N} \left\| \mathbf{x}_i - \mathbf{c}_{\text{label}(i)} \right\|^2
\]
where \( \mathbf{c}_{\text{label}(i)} \) denotes the centroid of the cluster to which point \( \mathbf{x}_i \) is assigned. Lower CP values indicate tighter, more cohesive clusters.

\paragraph{Dunn Validity Index (DVI)}  
The Dunn Index evaluates the ratio between the minimum inter-cluster distance and the maximum intra-cluster diameter \cite{dunn_1974,bezdek1998some}:
\[
\text{DVI} = \frac{ \displaystyle \min_{i \neq j} \delta(C_i, C_j) }{ \displaystyle \max_{k} \Delta(C_k) }
\]
where:
\begin{itemize}
    \item \( \delta(C_i, C_j) \) is the minimum pairwise distance between points in clusters \( C_i \) and \( C_j \),
    \item \( \Delta(C_k) \) is the maximum distance between any two points within cluster \( C_k \).
\end{itemize}
Higher DVI values indicate well-separated and compact clusters.

\paragraph{Silhouette Score}  
The silhouette score captures how similar an object is to its own cluster compared to other clusters. For each point \( \mathbf{x}_i \), the silhouette value \( s(i) \) is defined as \cite{batool2021clustering,rousseeuw_1987}:
\[
s(i) = \frac{b(i) - a(i)}{\max\{a(i), b(i)\}}
\]
where:
\begin{itemize}
    \item \( a(i) \) is the average intra-cluster distance (cohesion),
    \item \( b(i) \) is the lowest average inter-cluster distance to any other cluster (separation).
\end{itemize}
The overall silhouette score is the mean of all \( s(i) \). Values close to 1 indicate good clustering; values near 0 suggest overlapping clusters.

\paragraph{Stability via Label Alignment}  
To assess the stability of clustering across multiple runs, we align cluster labels using the Hungarian matching algorithm applied to pairwise confusion matrices. The stability metric is defined as the average alignment accuracy over all pairwise label permutations \cite{fahad2014survey}:
\[
\text{Stability} = \frac{2}{R(R - 1)} \sum_{i < j} \text{Match}(R_i, R_j)
\]
where \( R \) is the number of clustering repetitions, and \( \text{Match}(R_i, R_j) \) is the fraction of matching labels between runs \( i \) and \( j \) after optimal alignment.\\

\textbf{Consensus clustering procedure}

To ensure robust and stable group identification in unlabeled data, we employed \textbf{consensus clustering} following the protocol of Monti et al.~\cite{monti2003consensus}.

\textbf{Step 1: Repeated clustering.} We ran the clustering algorithm \( R = 10 \) times on the same dataset using random initializations. Each run produced a partition \( P_r \), where \( r \in \{1, 2, ..., R\} \).

\textbf{Step 2: Consensus matrix construction.} An \( n \times n \) consensus matrix \( C \) was constructed as:

\[
C_{ij} = \frac{1}{R} \sum_{r=1}^{R} \mathbb{I}\left( x_i \text{ and } x_j \text{ belong to the same cluster in } P_r \right)
\]

where \( \mathbb{I} \) is the indicator function, and \( n \) is the number of data points.

\textbf{Step 3: Final clustering.} We computed a distance matrix \( D = 1 - C \) and applied \textit{hierarchical clustering with average linkage} (UPGMA) to assign the final cluster labels.

\textbf{Rationale.} Averaging performance metrics (e.g., RI, ARI) over multiple runs does not capture clustering stability. By contrast, consensus clustering records how frequently pairs of samples are assigned to the same cluster across repeated runs, thereby yielding a more reliable final grouping and providing both a visual and a quantitative measure of stability. In the present framework, this procedure is particularly useful because Qlustering exhibits relatively high run-to-run stability, so repeated runs tend to preserve a common underlying structure despite differences in initialization. Consensus clustering therefore serves not only as a generic post-processing step, but as a practical way to extract that shared structure and improve the robustness of the final assignment.\\
The underlying codes and training/validation datasets for this study are available in the code availability section.\\ 






\section*{Data availability}

The QM9 and Iris datasets analyzed in this study are publicly available from the sources cited in Refs.~\cite{ramakrishnan2014quantum,uci_iris}. The processed datasets used to generate the results reported here, together with the scripts required to reproduce the analysis and figures, are available in the code repository listed below.

\section*{Code availability}

The code used to generate the results reported in this study is publicly available at
\url{https://github.com/ShmueLorber/Qlustering}.
The repository includes the simulation, training, validation, and plotting scripts, together with the processed input data needed to reproduce the reported results. The code is released under the MIT License.

\section*{Acknowledgments}
This study received no external funding. We thank Dr. Ari Packman for insightful discussions that informed this work.

\section*{Author contributions}
S.L. and Y.D. conceived the study, developed the methodology, interpreted the results, and wrote the manuscript. S.L. performed the calculations and analyzed the data. All authors reviewed and approved the final manuscript.

\section*{Competing interests}
The authors declare no competing interests.

\appendix

\section{Normalization of the Iris dataset}
\label{appendix: iris}
\begin{figure*}[ht]
      \centering

    \subfloat[]{\includegraphics[width=0.48\textwidth]{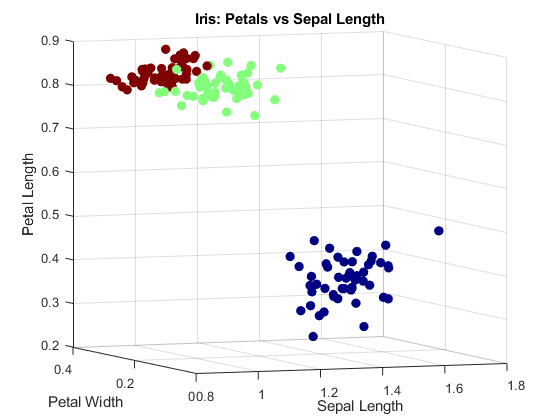}}
    \hfill
    \subfloat[]{\includegraphics[width=0.48\textwidth]{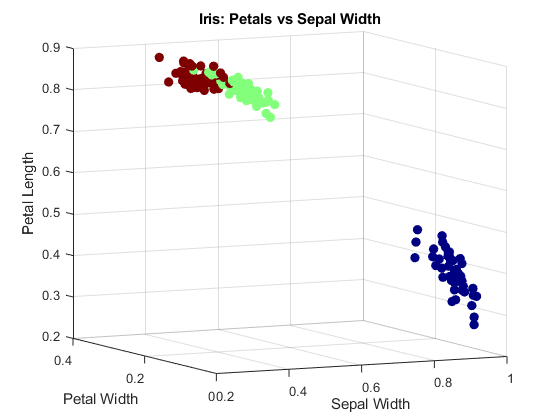}}

    \vspace{0.9em}

    \subfloat[]{\includegraphics[width=0.48\textwidth]{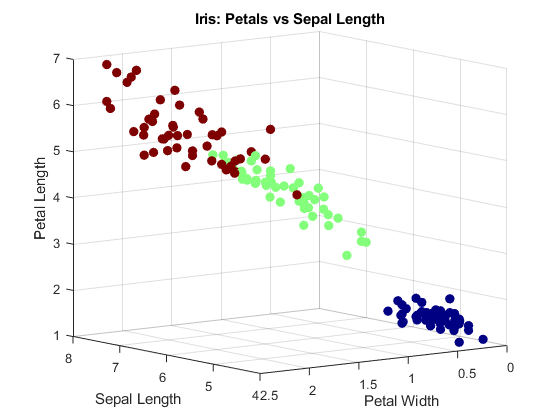}}
    \hfill
    \subfloat[]{\includegraphics[width=0.48\textwidth]{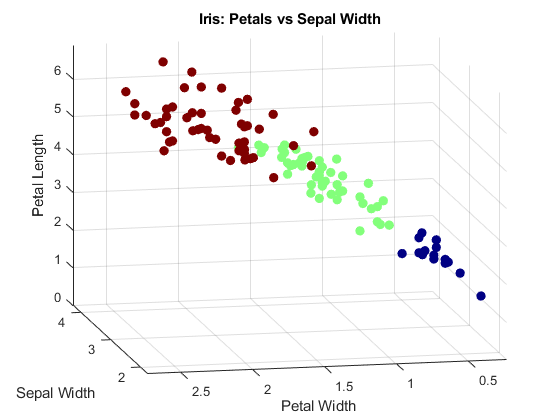}}

  \caption{\textbf{Feature distributions in the Iris dataset.}
  \textbf{(a)} Unnormalized, sepal length (less mixed);
  \textbf{(b)} Unnormalized, sepal width (mixed);
  \textbf{(c)} Normalized, sepal length (less mixed);
  \textbf{(d)} Normalized, sepal width (mixed). Notice how normalization skews the dataset, dispersing it across the field and bringing the different groups closer together}
  \label{fig:iris}
\end{figure*}

The effect of normalization is illustrated in Fig.~\ref{fig:iris}. Panels (a) and (b) display the Iris dataset before normalization, whereas panels (c) and (d) show the normalized data. The figure sheds light on a potential limitation of Qlustering—its reliance on normalization. Although a wide range of datasets can be embedded into Hilbert space using simple normalization, this step may distort the natural distribution of the points in the feature space.

In panels (a) and (b), the original Iris dataset is plotted in a three-dimensional representation (each plot traces one parameter to preserve the three-dimensional structure), showing clear separation between the blue group, \textit{Iris setosa}, and the other two overlapping groups, \textit{Iris versicolor} and \textit{Iris virginica}. Panels (c) and (d) present the normalized dataset in the same three-dimensional spaces. After normalization, the groups appear skewed and more widely spread, making them less separable and harder to detect.

This behavior can be detrimental when the original distribution already exhibits clear separation; however, in cases where the initial distribution is highly entangled, normalization may be beneficial. Furthermore, this limitation can be mitigated—for example, by introducing an ancilla parameter that preserves or encodes the initial spatial information.
\section{Robustness against dephasing}
\label{appendix: dephasing}
\begin{figure}[t]
    \centering
    \includegraphics[width=\linewidth]{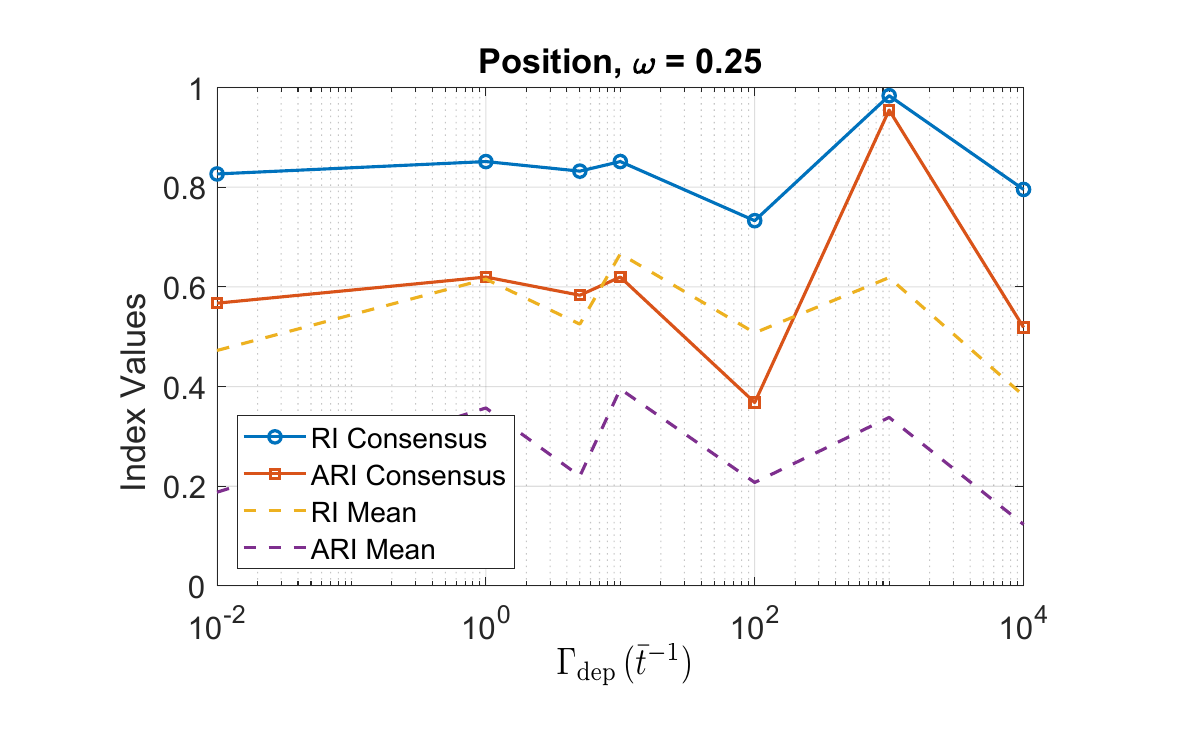}\\[0.25em]
    \includegraphics[width=\linewidth]{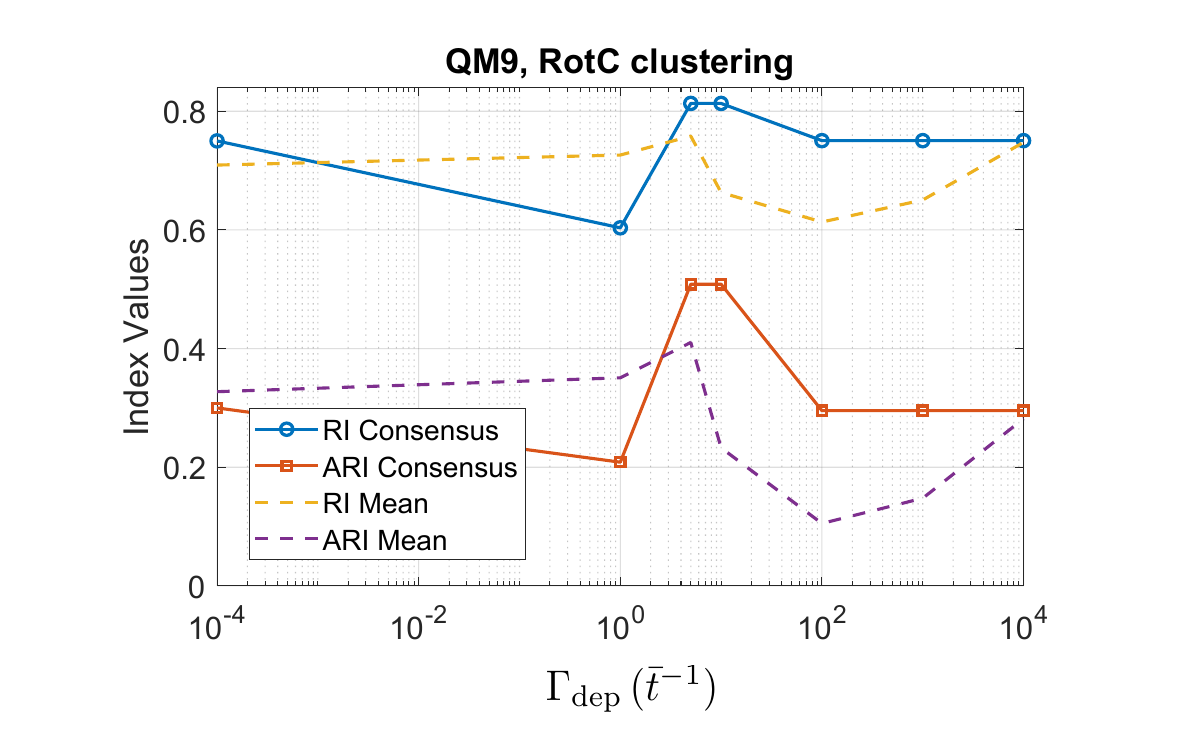}
    \caption{\textbf{Top: Effect of dephasing on the position-based task (Sec.~\ref{Position}, $\omega = 0.25$).} 
    Shown are RI and ARI values (mean and consensus) over ten Qlustering runs as a function of the dephasing rate $\Gamma_{\mathrm{dep}}$ in the range $0.01$--$10^4 \bar t^{-1}$. 
    60 state vectors divided into 4 groups were clustered. RI and ARI values remain relatively stable across this range, showing consistent performance and no systematic decline. 
    \textbf{Bottom: Dephasing robustness for a subset of the QM9 dataset containing 97 molecules.} 
    RI and ARI values (mean and consensus over ten Qlustering runs) were calculated using tagging based on the rotational constant $C$, with input vectors processed via SID; see Sec.~\ref{QM9} for further details about data preparation. These results demonstrate that Qlustering maintains reliable clustering even under strong dephasing.}
    \label{fig:dephasing}
\end{figure}

Adding dephasing to the network demonstrates clear robustness, as shown in Fig.~\ref{fig:dephasing} top. Dephasing was implemented using Zeno-type local dephasing terms in the Lindbladian~\cite{Breuer2002}, with the $V$-operators defined as $V_j = (\Gamma_{\mathrm{dep}})^{1/2} c^\dagger_j c_j$, where $j$ runs over all network sites. Here, $\Gamma_{\mathrm{dep}}$ denotes the dephasing rate expressed in natural (Hamiltonian) units of $\bar{t}^{-1}$, corresponding to the hopping amplitude $h$ in the Hamiltonian (see Eq.~\ref{eq: Hamiltonian}).

In Fig.~\ref{fig:dephasing} top, values of $\Gamma_{\mathrm{dep}}$ between $0.01$ and $1\cdot10^{4} t^{-1}$ were applied to the networks, followed by ten consecutive Qlustering runs on two of the tasks above. The first is the position-based task described in Sec.~\ref{Position}, with $\omega = 0.25$, $L = 3$, and $q = 4$, using the base points
$\{0.99, 0.11, 0.11\}$,
$\{0.11, 0.99, 0.11\}$,
$\{0.11, 0.11, 0.99\}$, and
$\tfrac{1}{\sqrt{3}}\{1,1,1\}$ (Fig.~\ref{fig:dephasing} top).  
The second is the real-world QM9 task, using the same subset of 97 molecules as in Sec.~\ref{QM9}. Evaluation of RI and ARI was carried out using both mean and consensus values, with labels based on the rotational constant C; the results are shown in Fig.~\ref{fig:dephasing} bottom. 

Note how the RI and ARI values fluctuate with $\Gamma_{\mathrm{dep}}$, showing a slight improvement at higher dephasing values. This robustness is a major advantage of Qlustering for implementation (most importantly in a general-purpose quantum computer), as it can operate reliably across various materials and environments, maintaining stable evaluation without the need for error correction or concern for noise-induced data corruption.
For reference, the typical hopping terms for transmon superconducting qubits are characterized by the Josephson energy $E_J \approx 1$–$10\,\text{GHz}$~\cite{krantz2019quantum}, which dominates the transmon Hamiltonian. This implies that $\bar{t}^{-1} \approx E_J \approx 1$–$10\,\text{GHz}$, so the characteristic hopping time is $\bar{t} \approx 0.1$–$1\,\text{ns}$. In our simulations, dephasing times can preserve Qlustering behavior even down to $\tau_{\mathrm{dep}} \approx 10^{-5} \bar{t} \approx 10^{-15}\,\text{s}$ for $\bar{t} = 0.1\,\text{ns}$.

The same dephasing calculations were performed for the other two tasks presented in the paper, yielding similar results.\\

\bibliographystyle{quantum}
\bibliography{references_with_doi}
\end{document}